\documentclass[twocolumn,showpacs,showkeys,preprintnumbers,amssymb,aps,superscriptaddress,prb]{revtex4}
\usepackage{graphicx}
\usepackage{dcolumn}
\usepackage{color}
\usepackage{bm}

\begin{document}


\title{Ising versus Potts criticality in a low-temperature magnetothermodynamics of a frustrated spin-1/2 Heisenberg triangular bilayer}
\author{Jozef Stre\v{c}ka}
\email{jozef.strecka@upjs.sk}
\affiliation{Institute of Physics, Faculty of Science, P. J. \v{S}af\'{a}rik University, Park Angelinum 9, 04001 Ko\v{s}ice, Slovakia}
\author{Katar\'ina Kar\v{l}ov\'a}
\affiliation{Institute of Physics, Faculty of Science, P. J. \v{S}af\'{a}rik University, Park Angelinum 9, 04001 Ko\v{s}ice, Slovakia}
\author{Vasyl Baliha}
\affiliation{Institute for Condensed Matter Physics, NASU, Svientsitskii Street 1, 79011 L'viv, Ukraine}
\author{Oleg Derzhko}
\affiliation{Institute for Condensed Matter Physics, NASU, Svientsitskii Street 1, 79011 L'viv, Ukraine}
\affiliation{Department for Theoretical Physics, Ivan Franko National University of L'viv, Drahomanov Street 12, 79005 L'viv, Ukraine}

\date{\today}

\begin{abstract}
Low-temperature magnetization curves and thermodynamics of a frustrated spin-1/2 Heisenberg triangular bilayer with the antiferromagnetic intradimer interaction and either ferromagnetic or antiferromagnetic interdimer interaction are investigated in a highly frustrated parameter region, where localized many-magnon eigenstates provide the most dominant contribution to  magnetothermodynamics. Low-energy states of the highly frustrated spin-1/2 Heisenberg triangular bilayer can be accordingly found from a mapping correspondence with an effective triangular-lattice spin-1/2 Ising model in a field. A description based on the effective Ising model implies that the frustrated Heisenberg triangular bilayer with the ferromagnetic interdimer coupling displays in a zero-temperature magnetization curve discontinuous magnetization jump, which is reduced upon increasing of temperature until a continuous field-driven phase transition from the Ising universality class is reached at a certain critical temperature. The frustrated Heisenberg triangular bilayer with the antiferromagnetic interdimer coupling contrarily exhibits multistep magnetization curve with intermediate plateaus at one-third and two-thirds of the saturation magnetization, whereas discontinuous magnetization jumps observable at zero temperature change to continuous field-driven phase transitions from the universality class of three-state Potts model at sufficiently low temperatures. Exact results and Monte Carlo simulations of the effective Ising model are confronted with full exact diagonalization data for the Heisenberg triangular bilayer in order to corroborate these findings. 
\end{abstract}

\maketitle

\section{Introduction}

The theorem due to Mermin and Wagner \cite{merm66} claims that temperature-driven phase transitions associated with a spontaneous breaking of the continuous symmetry of the isotropic Heisenberg model can be excluded in low spatial dimensions one and two on assumption that an external magnetic field is absent. However, temperature-driven phase transitions of the low-dimensional Heisenberg model cannot be definitely ruled out in presence of the magnetic field, because the magnetic field breaks a time-reversal symmetry and Mermin-Wagner theorem is inapplicable.\cite{merm66} From this perspective, the isotropic Heisenberg model often displays in presence of the magnetic field much greater diversity of classical and quantum phase transitions than its zero-field counterpart.  

Zero- and low-temperature magnetization curves of the isotropic Heisenberg model on several low-dimensional lattices may thus involve 
a lot of unconventional features,\cite{lacr11} which come from field- or temperature-driven phase transitions closely connected with appearance of fractional magnetization plateaus,\cite{hone04,meis07} magnetization jumps,\cite{schu02,shap02} magnetizations ramps \cite{naka10,saka11} or quantum spin-liquid states.\cite{lhui02,bale10,bale17} 

Over the past few years a great deal of attention has been paid to the Heisenberg bilayers, which exhibit a great variety of quantum phases and phase transitions.\cite{wang06,balc09,oitm12,szal12,szal13,zhan14,balc14,helm14,deva14,loho15,gome16,zhan16,bish17,bili17,souz18,stap18} 
The frustrated spin-$\frac{1}{2}$ Heisenberg square bilayer for instance displays a peculiar zero-field ground-state phase diagram including two lines of discontinuous and continuous phase transitions, which meet together at a peculiar quantum critical end point.\cite{stap18} Moreover, the frustrated spin-$\frac{1}{2}$ Heisenberg square\cite{rich06,derz11,derz08,derz10,karl18} and honeycomb\cite{krok17,krok18} bilayers belong to a valuable class of frustrated quantum spin systems, which exhibit a magnon-crystal state manifested in zero- and low-temperature magnetization curves as the last intermediate plateau emergent below the saturation field. The magnon-crystal phase is in its essence localized many-magnon eigenstate, which can be comprehensively described through a classical lattice-gas model or an equivalent Ising model within the framework of the localized-magnon approach (see Refs. \onlinecite{zhit05,derz06,derz15} and references therein).

In the present work we will investigate in detail low-temperature magnetization curves and thermodynamics of the quantum spin-$\frac{1}{2}$ Heisenberg triangular bilayer, which will display outstanding critical points falling either into the universality class of two-dimensional Ising model or two-dimensional three-state Potts model. It will be shown hereafter that the type of critical behavior depends fundamentally upon character of the interdimer interaction. To verify this conjecture, we will take advantage of several powerful analytical and numerical methods such as the variational technique, the localized-magnon approach, the many-body perturbation theory, the exact diagonalization and classical Monte Carlo simulations. 

The organization of this paper is as follows. The quantum spin-$\frac{1}{2}$ Heisenberg triangular bilayer is introduced in Sec. \ref{sec:model} along with basic steps of analytical and numerical methods used for its treatment. The most interesting results for the magnetization process and low-temperature thermodynamics are discussed in Sec. \ref{sec:result}. Finally, several concluding remarks and future outlooks are mentioned in Sec. \ref{sec:conc}. Some lenghtly calculations are put into two appendixes.

\section{Heisenberg triangular bilayer}
\label{sec:model}

\begin{figure}
\begin{center}
\includegraphics[width=0.5\columnwidth]{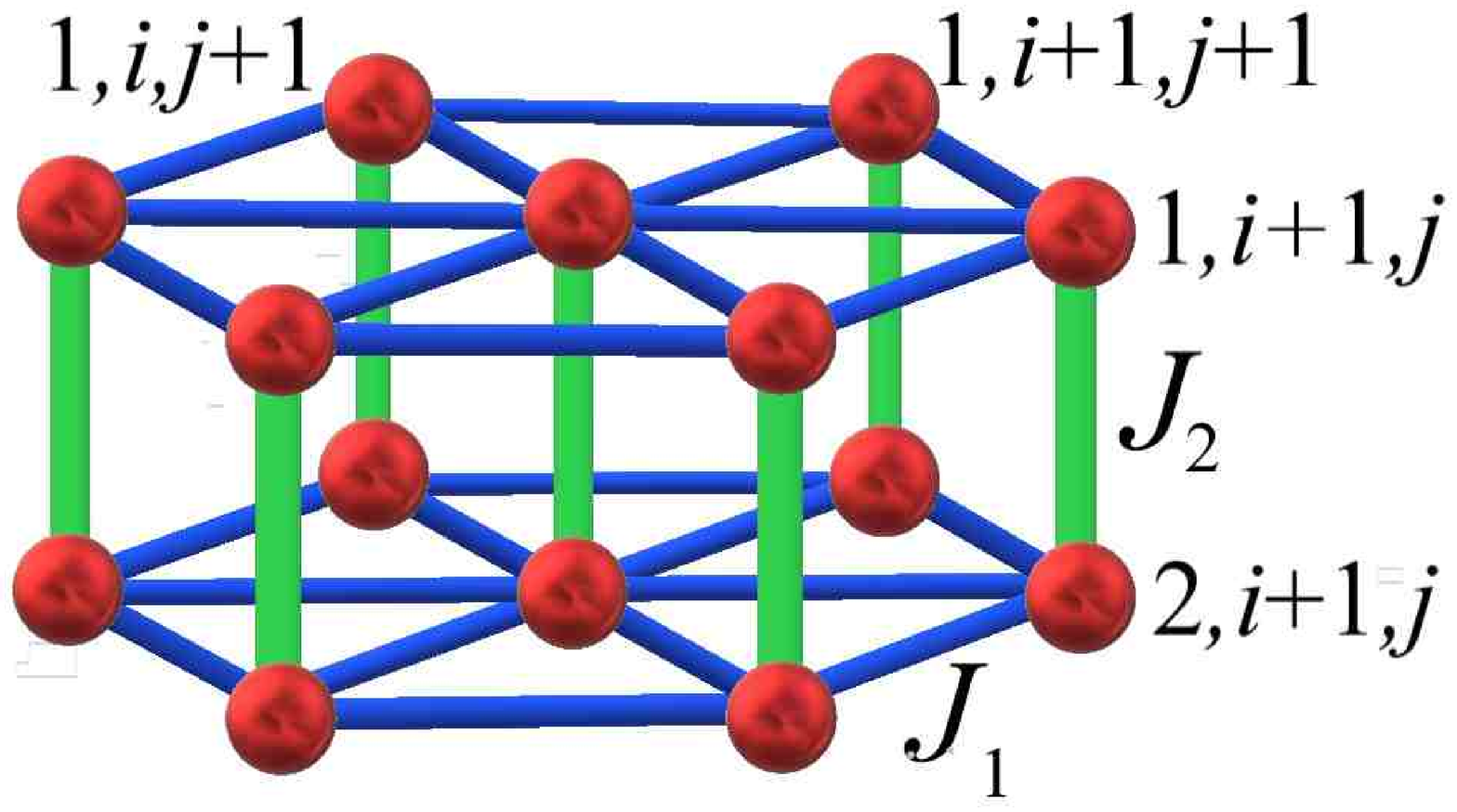}
\hspace*{-0.2cm}
\includegraphics[width=0.5\columnwidth]{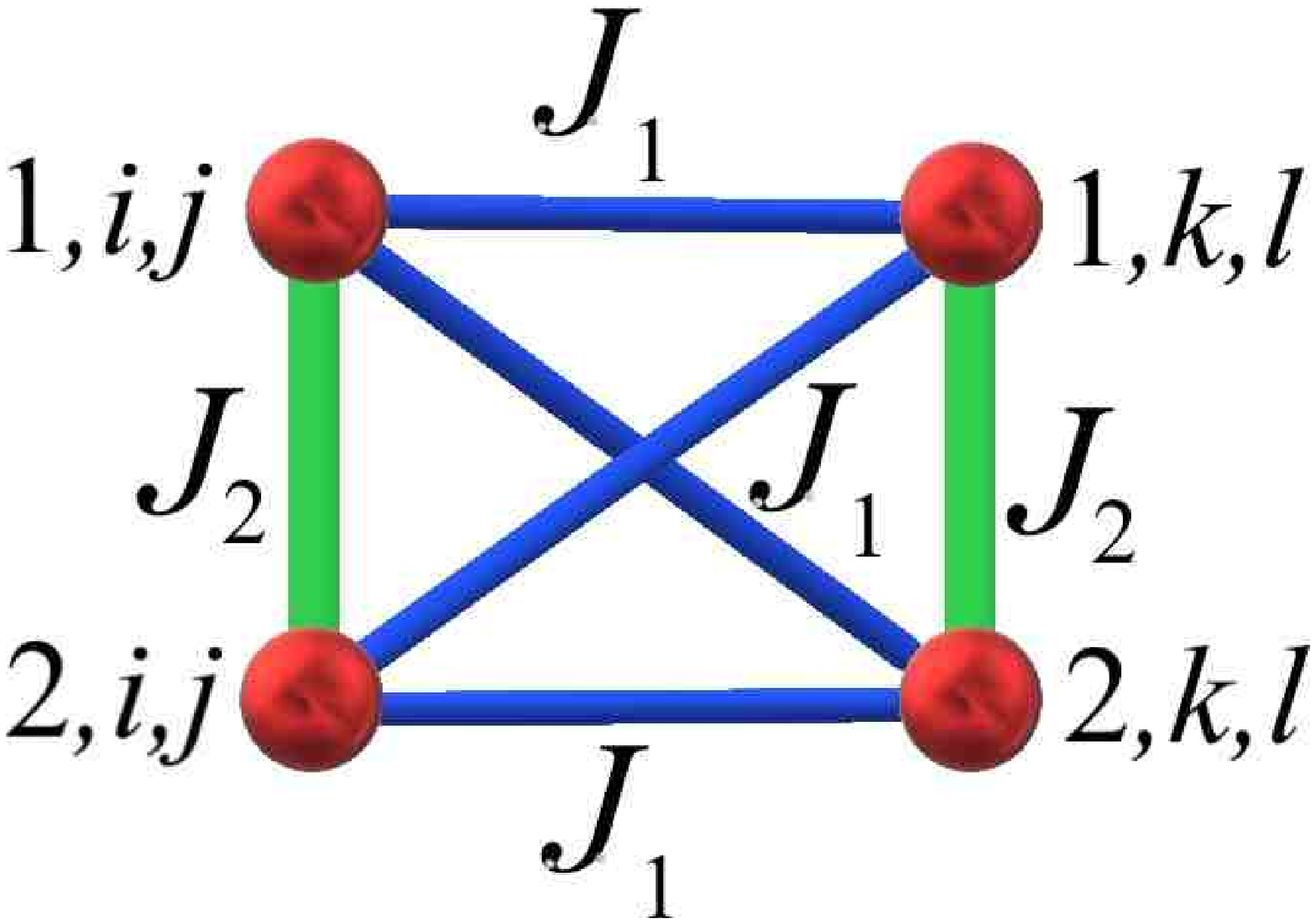}
\end{center}
\vspace{-0.8cm}
\caption{Left:  A small segment from the triangular bilayer. Thick (green) lines represent the intradimer coupling $J_2$, while thin (blue) lines correspond to the interdimer coupling $J_1$ within individual triangular layers. The interdimer couplings $J_1$ between the next-nearest-neighbor spins from different layers are not drawn for clarity. Right: A schematic illustration of all interaction terms of two neighboring spin dimers forming an elementary square plaquette.}
\label{fig1}
\end{figure}

Let us consider the frustrated spin-$\frac{1}{2}$ Heisenberg triangular bilayer (see a schematic illustration depicted on the left of Fig.~\ref{fig1}) defined through the Hamiltonian:
\begin{eqnarray}
\label{ham}
\hat{\cal H} \!\!\! &=& \!\!\! J_1 \!\!\!
\sum_{i,j=1}^{L} \! \sum_{l=1}^{2} \! (\boldsymbol{\hat{S}}_{1,i,j} \!\!+\!\! \boldsymbol{\hat{S}}_{2,i,j}) \!\cdot\! 
(\boldsymbol{\hat{S}}_{l,i+1,j} \!\!+\!\! \boldsymbol{\hat{S}}_{l,i,j+1} \!\!+\!\! \boldsymbol{\hat{S}}_{l,i+1,j+1}) \nonumber \\
\!\!\!&+&\!\!\! J_2 \sum_{i,j=1}^{L}  \boldsymbol{\hat{S}}_{1,i,j} \!\cdot\! \boldsymbol{\hat{S}}_{2,i,j} 
- h \sum_{i,j=1}^{L} \sum_{l=1}^{2}\hat{S}_{l,i,j}^z,
\end{eqnarray}
where $\boldsymbol{\hat{S}}_{l,i,j} \equiv (\hat{S}_{l,i,j}^x, \hat{S}_{l,i,j}^y, \hat{S}_{l,i,j}^z)$ denotes a spin-$\frac{1}{2}$ operator placed at a lattice site unambiguously determined by three subscripts. The first subscript $l=1,2$ determines a triangular layer, while the second (third) subscript specifies row (column) within a given layer (see the left panel in Fig.~\ref{fig1}). The coupling constant $J_1$ labels the Heisenberg interdimer interaction between the nearest-neighbor spins within each triangular layer (see thin blue lines in the left panel of Fig.~\ref{fig1}) and the next-nearest-neighbor spins from different triangular layers (not drawn in the left panel of Fig.~\ref{fig1}). The coupling constant $J_2$ labels the Heisenberg intradimer interaction between the nearest-neighbor spins from different layers and finally, the Zeeman's term $h \geq 0$ accounts for a magnetostatic energy of magnetic moments in an external magnetic field. In all subsequent calculations we will consider a triangular bilayer with the linear size $L$ and the total number of vertical dimers $N=L^2$ (i.e. the total number of spins $2N$) by imposing the periodic boundary conditions for  convenience. The Hamiltonian (\ref{ham}) can be solved by making use of several complementary analytical and numerical approaches, which will be dealt with in what follows. 

\subsection{Variational method}
The frustrated spin-$\frac{1}{2}$ Heisenberg triangular bilayer may exhibit in a highly frustrated parameter region $J_2 \gg |J_1|$ a \textit{singlet-dimer ground state}: 
\begin{eqnarray}
| SD \rangle = \prod_{i,j=1}^L \frac{1}{\sqrt{2}} \left(|\!\uparrow \rangle_{1,i,j} |\!\downarrow \rangle_{2,i,j} - |\!\downarrow \rangle_{1,i,j} |\!\uparrow \rangle_{2,i,j} \right),
\label{sgs}
\end{eqnarray}
which is constituted by a product of singlet states formed between the nearest-neighbor spins from adjacent layers. A rigorous criterion for appearance of the singlet-dimer ground state (\ref{sgs}) can be found through the variational principle.\cite{shas81,bose92} To this end, the total Hamiltonian (\ref{ham}) of the frustrated spin-$\frac{1}{2}$ Heisenberg triangular bilayer can be first decomposed into the local Hamiltonians of square sub-units $\hat{\cal H} = \sum_{t=1}^{3N} \hat{\cal H}_t$, whereas each local Hamiltonian of a square sub-unit involves all the interaction terms of two nearest-neighbor spin dimers: 
\begin{eqnarray}
\hat{\cal H}_t \!\!\!&=&\!\!\! J_1 (\boldsymbol{\hat{S}}_{1,i,j} + \boldsymbol{\hat{S}}_{2,i,j}) \!\cdot\! (\boldsymbol{\hat{S}}_{1,k,l} + \boldsymbol{\hat{S}}_{2,k,l}) \nonumber \\
               \!\!\!&+&\!\!\! \frac{J_2}{6} (\boldsymbol{\hat{S}}_{1,i,j} \!\cdot\! \boldsymbol{\hat{S}}_{2,i,j} + \boldsymbol{\hat{S}}_{1,k,l} \!\cdot\! \boldsymbol{\hat{S}}_{2,k,l}) \nonumber \\
							 \!\!\!&-&\!\!\! \frac{h}{6} ({\hat{S}}_{1,i,j}^z + {\hat{S}}_{2,i,j}^z + {\hat{S}}_{1,k,l}^z + {\hat{S}}_{2,i,j}^z),
\label{th}
\end{eqnarray}
see the right panel in Fig. \ref{fig1}. The factor $1/6$ at the intradimer interaction $J_2$ and the magnetic-field term $h$ avoids overcounting of these interactions terms, which are symmetrically split into six different local Hamiltonians of square sub-units. The variational procedure then provides the lower bound for the ground-state energy of the frustrated spin-$\frac{1}{2}$ Heisenberg triangular bilayer:
\begin{eqnarray}
E_0 = \langle \Psi_0 | \hat{\cal H} | \Psi_0 \rangle = \langle \Psi_0 | \sum_{t=1}^{3N} \! \hat{\cal H}_{t} | \Psi_0 \rangle 
\!\geq\! \sum_{t=1}^{3N} \varepsilon_{t}^0,
\label{var}
\end{eqnarray}
because the ground-state energy $E_0$ corresponding to the eigenvector $| \Psi_0 \rangle$ must be necessarily greater than or equal to the sum of the lowest-energy eigenvalues of the square sub-units $\varepsilon_{t}^0$. The eigenenergies of the spin-$\frac{1}{2}$ Heisenberg square with the coupling constants $J_1$ and $J_2/6$ defined by Eq. (\ref{th}) can be expressed in terms of four quantum spin numbers $S_t$, $S_{i,j}$, $S_{k,l}$ and $S_t^z$:
\begin{eqnarray}
\varepsilon_{t} \!\!\!&=&\!\!\! \frac{J_1}{2} [S_{t}(S_{t}+1) - S_{i,j} (S_{i,j} + 1) - S_{k,l} (S_{k,l} + 1)]  \nonumber \\ 
\!\!\!&+&\!\!\! \frac{J_2}{12} [S_{i,j} (S_{i,j} + 1) + S_{k,l} (S_{k,l} + 1)] - \frac{J_2}{4} - \frac{h}{6} S_t^z,
\label{spec}
\end{eqnarray}
which determine eigenvalues for the total spin of two nearest-neighbor spin pairs $\boldsymbol{\hat{S}}_{i,j} = \boldsymbol{\hat{S}}_{1,i,j} + \boldsymbol{\hat{S}}_{2,i,j}$ and $\boldsymbol{\hat{S}}_{k,l}= \boldsymbol{\hat{S}}_{1,k,l} + \boldsymbol{\hat{S}}_{2,k,l}$ coupled through the intradimer interaction $J_2$, the total spin of a square sub-unit $\boldsymbol{\hat{S}}_{t} = \boldsymbol{\hat{S}}_{i,j} + \boldsymbol{\hat{S}}_{k,l}$ and its $z$-component ${\hat{S}}_{t}^z = {\hat{S}}_{i,j}^z + {\hat{S}}_{k,l}^z$, respectively. The lowest-energy eigenvalues of the spin-$\frac{1}{2}$ Heisenberg square sub-unit are listed below for admissible combinations of quantum spin numbers in the following order 
$\varepsilon_{t} \left(S_t, S_{ij}, S_{kl}, S_t^z \right)$:
\begin{eqnarray}
\varepsilon_{t} \left(0, 0, 0, 0 \right) \!\!\!&=&\!\!\! - \frac{J_2}{4}, \label{speca} \\
\varepsilon_{t} \left(1, 0, 1, 1 \right) \!\!\!&=&\!\!\! \varepsilon_{t} \left(1, 1, 0, 1 \right) = - \frac{J_2}{12} - \frac{h}{6}, \label{specb} \\
\varepsilon_{t} \left(0, 1, 1, 0 \right) \!\!\!&=&\!\!\! -2J_1 + \frac{J_2}{12}, \label{specc} \\
\varepsilon_{t} \left(1, 1, 1, 1 \right) \!\!\!&=&\!\!\! -J_1 + \frac{J_2}{12} - \frac{h}{6}, \label{specd} \\
\varepsilon_{t} \left(2, 1, 1, 2 \right) \!\!\!&=&\!\!\! J_1 + \frac{J_2}{12} - \frac{h}{3}. \label{spece}
\end{eqnarray} 
It is worthwhile to remark that the lower bound for the ground-state energy obtained from the eigenenergy (\ref{speca}) coincides with the energy of the singlet-dimer phase (\ref{sgs}), which consequently represents the true ground state of the frustrated spin-$\frac{1}{2}$ Heisenberg triangular bilayer whenever the eigenenergy (\ref{speca}) is lower than all the other ones (\ref{specb})-(\ref{spece}). If one considers the spin-$\frac{1}{2}$ Heisenberg triangular bilayer with the ferromagnetic interdimer coupling $J_1<0$ to be further referred to as FM/AF bilayer one consequently obtains the following sufficient condition $J_2 > 3|J_1|$, $h < J_2 - 3|J_1|$ for the appearance of the singlet-dimer ground state (\ref{sgs}). On the other hand, the sufficient condition for emergence of the singlet-dimer ground state (\ref{sgs}) is shifted to a more frustrated parameter space $J_2 > 6J_1$, $h < J_2$ for the spin-$\frac{1}{2}$ Heisenberg triangular bilayer with the antiferromagnetic interdimer coupling $J_1>0$ to be further referred to as AF/AF bilayer. 

\subsection{Exact one-magnon eigenstates}

Exact eigenstates of the frustrated spin-$\frac{1}{2}$ Heisenberg triangular bilayer can be rigorously obtained within one-magnon subspace with the $z$-component of the total spin $S_{T}^z = N - 1$. If the energy of one-magnon eigenstates $E_{\bf{k}}^{(j)}$ ($j=1,2$)  is quoted relative to the energy of fully polarized ferromagnetic state $E_{\rm FM} = N(\frac{1}{4} J_2 + 3 J_1 -h)$, then, one obtains after diagonalization of the Hamiltonian (\ref{ham}) in the one-magnon subspace (see Appendix \ref{appa}) the following relative eigenenergies 
$\varepsilon_{\bf{k}}^{(j)} = E_{\bf{k}}^{(j)} - E_{\rm FM}$:
\begin{eqnarray}
\varepsilon_{\bf{k}}^{(1)} \!\!\!&=&\!\!\! -J_2 - 6J_1 + h, \label{oma} \\
\varepsilon_{\bf{k}}^{(2)} \!\!\!&=&\!\!\! 2J_1\left[\cos k_a+\cos k_b+\cos\left(k_a+k_b\right)-3\right] + h \nonumber \\
\!\!\!&=&\!\!\! 8J_1\left(\cos \frac{k_a}{2}\cos \frac{k_b}{2}\cos\frac{k_a+k_b}{2}-1 \right) + h. \label{omb}
\end{eqnarray}
Here $k_a=k_x$ and $k_b=-k_x/2+\sqrt{3}k_y/2$ (the triangle side length $a_0=1$). It is quite obvious that the one-magnon energy spectrum of the frustrated spin-$\frac{1}{2}$ Heisenberg triangular bilayer consists of two energy bands, whereas the former band (\ref{oma}) is completely flat (dispersionless) in opposite to the latter dispersive band (\ref{omb}). 
\begin{figure}
\begin{center}
\includegraphics[width=0.5\textwidth]{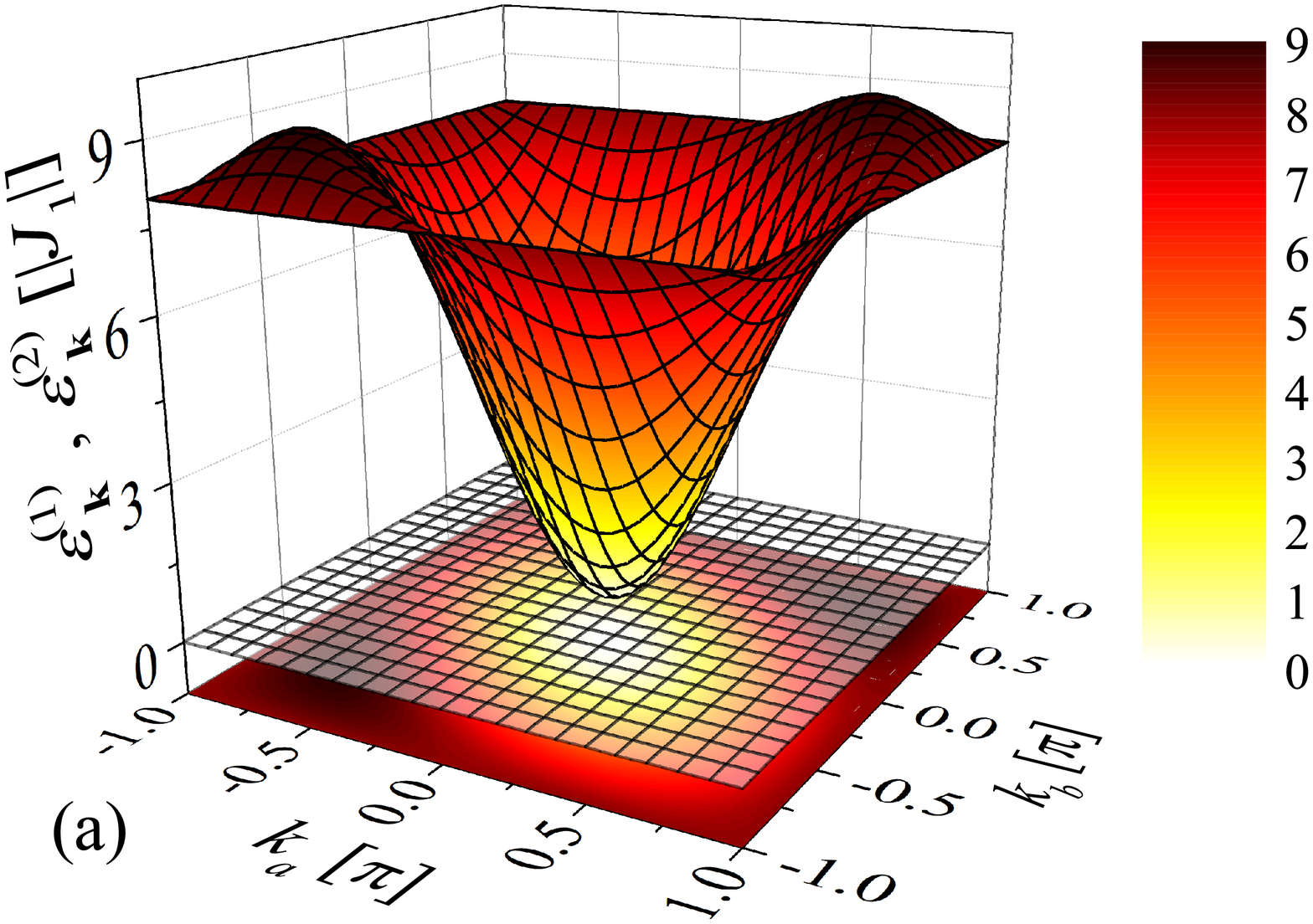}
\includegraphics[width=0.5\textwidth]{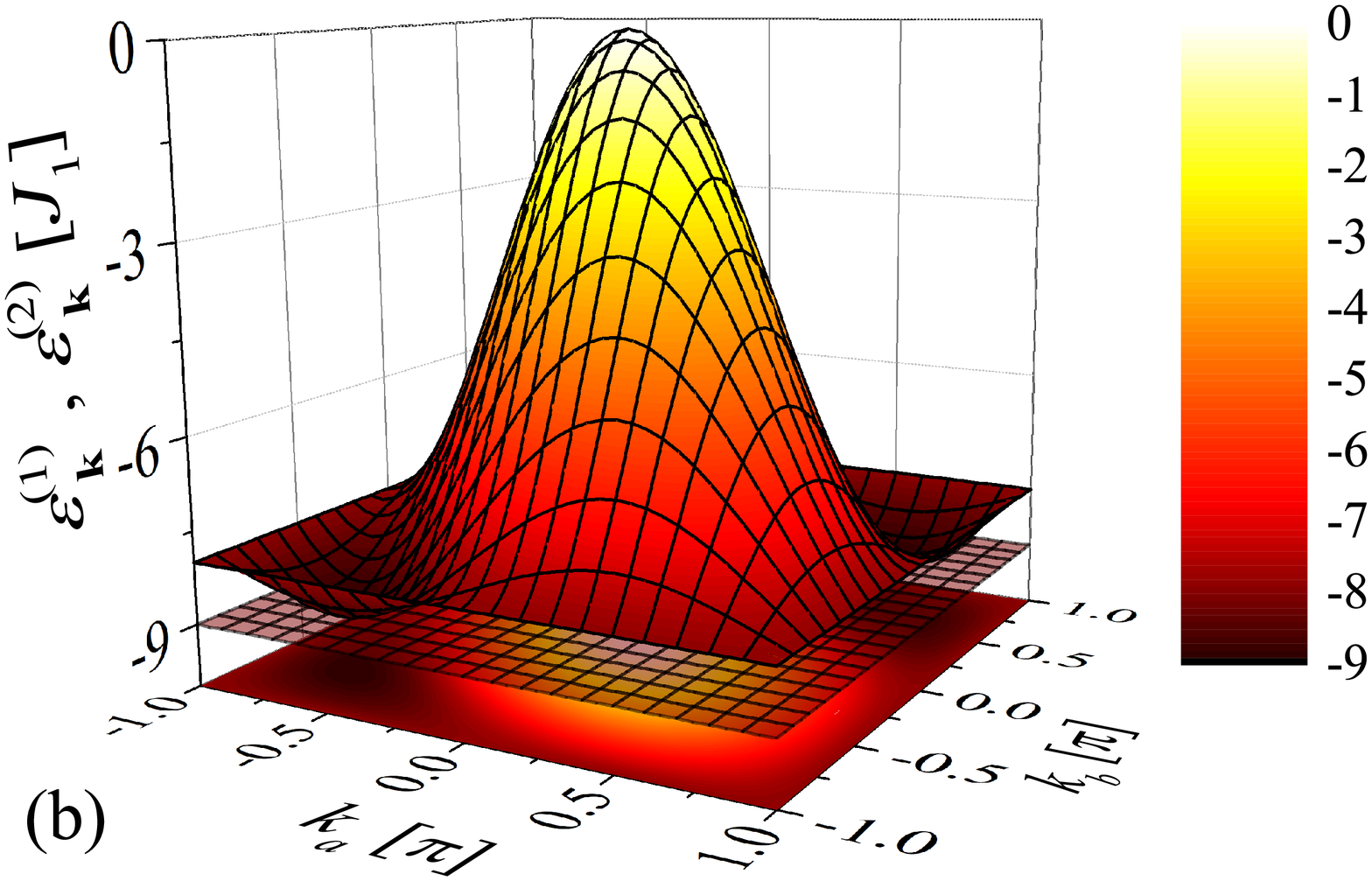}
\end{center}
\vspace{-1.2cm}
\caption{One-magnon bands of the spin-$\frac{1}{2}$ Heisenberg triangular bilayer by considering zero magnetic field 
and: (a) ferromagnetic interdimer interaction $J_1<0$ and the relative ratio $J_2/|J_1|=6$; (b) antiferromagnetic interdimer interaction $J_1>0$ and the relative ratio $J_2/J_1=3$. A projection of the dispersive band (\ref{omb}) into $k_a-k_b$ plane is also shown as a contour plot, while the interaction ratio was chosen for two particular values when the flat band (\ref{oma}) touches the lowest energy of the dispersive band (\ref{omb}).}
\label{fig2}
\end{figure}
It could be easily checked that the flat band with the relative eigenenergy (\ref{oma}) corresponds to a singlet-dimer state: 
\begin{eqnarray}
| s \rangle_{i,j} = \frac{1}{\sqrt{2}} \left(|\!\uparrow \rangle_{1,i,j} |\!\downarrow \rangle_{2,i,j} - |\!\downarrow \rangle_{1,i,j} |\!\uparrow \rangle_{2,i,j} \right),
\label{sd}
\end{eqnarray}
which represents localized one-magnon state at one vertical dimer connected through the intradimer interaction $J_2$. The dispersive energy band (\ref{omb}) gives the lowest energy $\varepsilon_{\bf{k}, {\rm min}}^{(2)}=h$ at $k_a=k_b=0$ for the frustrated FM/AF bilayer with the ferromagnetic interdimer coupling $J_1<0$ and $\varepsilon_{\bf{k}, {\rm min}}^{(2)}=-9J_1+h$ at $k_a=k_b= \pm 2\pi/3$ for the frustrated AF/AF bilayer with the antiferromagnetic interdimer coupling $J_1>0$. Owing to this fact, the flat band becomes the lowest energy one-magnon eigenstate in the parameter region $J_2>6|J_1|$ for the FM/AF bilayer with $J_1<0$ and $J_2>3J_1$ for the AF/AF bilayer with $J_1>0$ as displayed in the upper and lower panels of Fig. \ref{fig2}, respectively.

\subsection{Localized many-magnon eigenstates}

Localized many-magnon eigenstates of the frustrated spin-$\frac{1}{2}$ Heisenberg triangular bilayer can be obtained from the fully polarized FM state by populating localized one-magnon (singlet-dimer) state (\ref{sd}) on some of its vertical dimers. Such a construction of exact many-magnon eigenstates relies on the fact that the total spin of the vertical dimers represents a conserved quantity, because the corresponding spin operator  $\boldsymbol{\hat{S}}_{i,j}= \boldsymbol{\hat{S}}_{1,i,j} + \boldsymbol{\hat{S}}_{2,i,j}$ commutes with the Hamiltonian (\ref{ham}). It is therefore quite convenient to reexpress the Hamiltonian (\ref{ham}) of the frustrated spin-$\frac{1}{2}$ Heisenberg triangular bilayer in terms of the total spin of vertical dimers:
\begin{eqnarray}
\label{hamdb}
\hat{\cal H} \!\!\! &=& \!\!\! J_1 \!
\sum_{i,j=1}^{L} \! \boldsymbol{\hat{S}}_{i,j} \!\cdot\! (\boldsymbol{\hat{S}}_{i+1,j} \!+\! \boldsymbol{\hat{S}}_{i,j+1} \!+\! \boldsymbol{\hat{S}}_{i+1,j+1}) \nonumber \\
\!\!\!&+&\!\!\! \frac{1}{2} J_2 \sum_{i,j=1}^{L} \boldsymbol{\hat{S}}_{i,j}^2 - h \sum_{i,j=1}^{L} \hat{S}_{i,j}^z - \frac{3}{4} N J_2,
\end{eqnarray}
which directly implies that flat excitation bands can be constructed from the fully polarized FM state by making use of the bound one-magnon (singlet-dimer) state (\ref{sd}). It is of principal importance that a spin pair in the singlet state (\ref{sd}) is effectively decoupled from other spins and hence, the completely flat many-magnon eigenstates can be constructed from the fully polarized FM state by an independent placing of the bound one-magnon state (\ref{sd}) on some of vertical dimers of a fully frustrated triangular bilayer. If two singlets are placed on the vertical dimers, which do not interact with each other through the interdimer interaction $J_1$, then, the overall energy is just a simple sum of eigenenergies of two independent localized one-magnon states (\ref{oma}). It should be realized, however, that an occupation of singlets on two adjacent vertical dimers causes a double counting of energy contributions arising from four common interdimer couplings $J_1$. Bearing all this in mind, the localized many-magnon eigenstates of the frustrated spin-$\frac{1}{2}$ Heisenberg triangular bilayer can be represented in the language of the classical lattice-gas model defined on a triangular lattice through the Hamiltonian:
\begin{eqnarray}
\label{hamlg}
{\cal H} \!\!\!&=&\!\!\! J_1 \sum_{i,j=1}^{L} (n_{i,j} n_{i+1,j} + n_{i,j} n_{i,j+1} + n_{i,j} n_{i+1,j+1})  \nonumber \\
\!\!\!&-&\!\!\! \mu \sum_{i,j=1}^{L} n_{i,j} + E_{\rm FM}.
\end{eqnarray}
The particular value of the occupation number $n_{i,j} = 1$ $(n_{i,j} = 0)$ corresponds to the singlet (polarized triplet) state and the chemical potential $\mu = J_2 + 6J_1 - h$ relates to an energy penalty associated with a creation of the singlet state (\ref{sd}) on a ferromagnetic background. The first term provides correction to an energy of independent localized one-magnon states, which is relevant for  two singlets placed on adjacent vertical dimers. 

For further convenience, it is advisable to pass from the effective lattice-gas model given by the  Hamiltonian (\ref{hamlg}) to the equivalent Ising model what can be achieved by relating the occupation number $n_{i,j} = 0,1$ with the two-valued Ising variable $\sigma_{i,j} = \pm 1$ through the transformation $n_{i,j} = (1 + \sigma_{i,j})/2$. The localized many-magnon eigenstates of the spin-$\frac{1}{2}$ Heisenberg triangular bilayer can be accordingly found from a mapping correspondence with the classical Ising model on a triangular lattice given by the effective Hamiltonian:
\begin{eqnarray}
\label{hamef}
{\cal H} \!\!\!&=&\!\!\! J_{\rm eff} \sum_{i,j=1}^{L} \!\! ({\sigma}_{i,j} {\sigma}_{i+1,j} + {\sigma}_{i,j} {\sigma}_{i,j+1} + {\sigma}_{i,j} {\sigma}_{i+1,j+1}) \nonumber \\
\!\!\!&-&\!\!\! h_{\rm eff} \sum_{i,j=1}^{L} {\sigma}_{i,j} + N \left(\frac{3}{4} J_1 - \frac{1}{4} J_2 - \frac{1}{2} h \right).
\end{eqnarray}
The parameters $J_{\rm eff}$ and $h_{\rm eff}$ represent the effective nearest-neighbor interaction and the effective field of the Ising model on a triangular lattice, whereas they are explicitly given by:
\begin{eqnarray}
\label{mp}
J_{\rm eff} = \frac{J_1}{4}, \qquad \qquad h_{\rm eff} = \frac{J_2 + 3 J_1 - h}{2}.
\end{eqnarray}
It is worthwhile to remark that the spin state $\sigma_{i,j} = +1$ ($\sigma_{i,j} = -1$) of the effective Ising model corresponds to a singlet (polarized triplet) state on a given vertical dimer. Hence, it follows that the magnetization of the spin-$\frac{1}{2}$ Heisenberg triangular bilayer can be calculated from the magnetization of the effective Ising model on a triangular lattice according to:
\begin{eqnarray}
\label{mag}
m \equiv \frac{1}{2} \langle \hat{S}_{1,i,j}^z \!+\! \hat{S}_{2,i,j}^z \rangle = \frac{1}{2} (1 \!-\! \langle n_{i,j} \rangle) = \frac{1}{4} (1 \!-\! \langle \sigma_{i,j} \rangle).
\end{eqnarray}

It is obvious that the Hamiltonian (\ref{hamef}) of the effective Ising model has Z$_2$ symmetry in contrast to SU(2)-symmetry of the isotropic Heisenberg model. This apparent contradiction reflects a fundamental property of the frustrated spin-$\frac{1}{2}$ Heisenberg triangular bilayer, which can be easily understood from the alternative reformulation of the investigated model system in the dimer basis (\ref{hamdb}). The singlet and polarized triplet states of each vertical dimer accordingly represent the only relevant states of the dimeric unit cell in a highly frustrated parameter region and this binary degree of freedom can be subsequently described by the Ising variable reflecting two different irreducible representations of the total spin of the dimeric unit cell. 

Moreover, the effective Hamiltonian given by Eqs.~(\ref{hamef}) and (\ref{mp}) may be obtained in a different manner by using the many-body perturbation theory when starting from a strong-coupling limit.\cite{fuld91,mila11} In fact, one may alternatively consider a set of $N$ non-interacting spin-$\frac{1}{2}$ Heisenberg dimers at the particular magnetic field $h_0=J_2$ when the energies of the polarized triplet state $|t\rangle_{i,j} =|\!\!\uparrow\rangle_{1,i,j}|\!\!\uparrow\rangle_{2,i,j}$ and the singlet (one-magnon) state $|s\rangle_{i,j}$ given by Eq. (\ref{sd}) coincide. The Hamiltonian of this main part is denoted as $\hat{{\cal{H}}}_0$. Treating the rest terms in the Hamiltonian (\ref{ham}) as a perturbation $\hat{{\cal{V}}}=\hat{{\cal{H}}}-\hat{{\cal{H}}}_0$, one may calculate the effective Hamiltonian according to the formula:\cite{fuld91} $\hat{{\cal {H}}}_{\rm eff}=P(\hat{{\cal {H}}}_0+\hat{{\cal{V}}})P+\ldots$, where $P=\prod_{i,j}(|t\rangle\langle t|+|s\rangle\langle s|)_{i,j}$ is the projector onto the $2^{N}$-fold degenerate space upon which the effective Hamiltonian $\hat{{\cal {H}}}_{\rm eff}$ acts. Introducing (pseudo)spin-$\frac{1}{2}$ operators $\hat{T}^z=(|t\rangle\langle t|-|s\rangle\langle s|)/2$, $\hat{T}^+=|t\rangle\langle s|$, and $\hat{T}^-=|s\rangle\langle t|$ for each spin-$\frac{1}{2}$ Heisenberg dimer given by site indices $i$ and $j$, one finds that $\hat{{\cal {H}}}_{\rm eff}$ is the effective Ising model on a triangular lattice defined through the Hamiltonian (\ref{hamef}) and (\ref{mp}) upon identifying $\sigma_{i,j}=-2\hat{T}^z_{i,j}$.

All basic magnetothermodynamic quantities of the effective Ising model on a triangular lattice given by Eqs. (\ref{hamef}) and (\ref{mp}) such as magnetization, susceptibility and specific heat can be obtained by exact calculations for small finite-size systems (see Appendix \ref{appb}) or by performing classical Monte Carlo (MC) simulations implementing standard Metropolis sampling for larger system sizes. In addition, a few exact results are known for the Ising model on a triangular lattice on assumption that the effective field or temperature becomes zero. Let us make a few implications arising from those rigorous results. It can be readily understood from Eq. (\ref{mp}) that the spin-$\frac{1}{2}$ Heisenberg FM/AF triangular bilayer with the ferromagnetic interdimer coupling $J_1<0$ is mapped onto the effective triangular Ising ferromagnet ($J_{\rm eff}<0$), which exhibits just two ground states with all spins being 'up' for $h_{\rm eff} > 0$ or all spins being 'down' for $h_{\rm eff} < 0$. The former ground state apparently corresponds to the singlet-dimer phase (\ref{sgs}), while the latter ground state corresponds to the fully polarized ferromagnetic phase. According to Eq. (\ref{mp}), the singlet-dimer ground state (\ref{sgs}) is favored before the ferromagnetic one at zero magnetic field $h=0$ for $J_2/|J_1|>3$ in concordance with the variational arguments. Moreover, the effective Ising triangular ferromagnet exhibits a continuous phase transition from the Ising universality class at the critical temperature $k_{\rm B} T_{c}/|J_{\rm eff}| = 4/ \ln 3$ on assumption that the effective field equals zero $h_{\rm eff} = 0$.\cite{hout50,temp50,wann50,domb60} This result would imply the Ising-type critical point (singularity) in the isothermal magnetization curve of the frustrated Heisenberg FM/AF triangular bilayer at the critical temperature $k_{\rm B} T_{c}/|J_1| = 1/\ln 3$ and the critical field $h_{c} = J_2 - 3|J_1|$ for $J_2/|J_1|>3$. In addition, the effective triangular Ising ferromagnet displays at low enough temperatures $T<T_c$ and zero effective field $h_{\rm eff} = 0$ a spontaneous long-range order, which is characterized by nonzero spontaneous magnetization acquiring two different values equal in magnitude but of opposite sign. A double solution for the spontaneous magnetization indicates for the spin-$\frac{1}{2}$ Heisenberg FM/AF triangular bilayer a phase coexistence due to a discontinuous field-induced phase transition at the critical field $h_{c} = J_2 - 3|J_1|$ if considering sufficiently low temperatures $T<T_{c}$. An exact result for the spontaneous magnetization of the triangular Ising ferromagnet \cite{pott52} thus enables a rigorous calculation of two magnetization values:
\begin{eqnarray}
\label{magpt}
m_{\pm} \!=\! \frac{1}{4} \!\! \left\{1 \!\pm\! \left[\!1 \!-\! \frac{16x^2}{(1+3x^2)(1-x^2)^3} \! \right]^{\!\frac{1}{8}} \!\right\}\!\!,\, 
x = {\rm e}^{-2 \beta J_{\rm eff}}
\end{eqnarray}
which determine a size of the magnetization jump at the discontinuous field-driven phase transition. 

On the other hand, the spin-$\frac{1}{2}$ Heisenberg AF/AF triangular bilayer with the antiferromagnetic interdimer coupling $J_1>0$ is mapped according to Eq. (\ref{mp}) onto the effective triangular Ising antiferromagnet ($J_{\rm eff}>0$), which displays four possible ground states with all spins being 'up', all spins being 'down', or with a period-three sequence of 'up-up-down' or 'up-down-down' states. Two former ground states repeatedly correspond to the singlet-dimer and the fully polarized ferromagnetic phases, while the latter two ground states correspond to a regular period-three alternation of the singlet and polarized triplet dimer states 'singlet-singlet-triplet' and 'singlet-triplet-triplet', respectively. 
All these ground states exhibit a spontaneous long-range order of spin states of the vertical dimers, whereas elementary excitation spectra can be straightforwardly obtained upon converting one singlet-dimer state to the polarized triplet state or vice versa.

It should be emphasized that the effective triangular Ising antiferromagnet does not display at zero effective field $h_{\rm eff} = 0$ the critical behavior from the Ising universality class due to a geometric spin frustration.\cite{wann50,domb60} However, it has been firmly established that the triangular Ising antiferromagnet shows outstanding criticality closely connected with a breakdown of the period-three 'up-up-down' (or 'up-down-down') ground state manifested in a low-temperature magnetization process as an intermediate one-third plateau, which disappears upon rising temperature at critical points from the universality class of three-state Potts model. At zero temperature, the saturation fields $h_{\rm eff}/ J_{\rm eff} = \pm 6$ of the effective triangular Ising antiferromagnet are thus consistent with the appearance and disappearance of the intermediate one-third and two-thirds magnetization plateaus of the Heisenberg AF/AF triangular bilayer. The one-third magnetization plateau should thus emerge at the critical field $h_{c1} = J_2$, while the two-thirds magnetization plateau should terminate at the critical field $h_{c3} = J_2 + 6J_1$. An abrupt magnetization jump associated with a field-driven phase transition between the intermediate one-third and two-thirds plateaus of the Heisenberg AF/AF triangular bilayer appears at zero effective field $h_{\rm eff} = 0$, which implies the following critical value of the magnetic field $h_{c2} = J_2 + 3J_1$. Note that the universality class of this phase transition at finite temperatures is still under debate, but there are strong indications of Kosterlitz-Thouless type phase transition.\cite{kinz81,nien84,qian04} It should be pointed out, moreover, that the magnetic behavior of the Heisenberg AF/AF triangular bilayer should be symmetric with respect to the critical field $h_{c2} = J_2 + 3J_1$, which represents zero effective field for the effective triangular Ising antiferromagnet.

\subsection{Mapping to hard-hexagon model} 
It is quite clear from Eq. (\ref{mp}) that the ferromagnetic interdimer coupling $J_1<0$ leads to an effective attraction between the singlets from neighboring dimers, while the antiferromagnetic interdimer coupling $J_1>0$ gives rise to an effective repulsion between the neighboring singlets. This latter observation would suggest that singlets residing on neighboring dimers should be forbidden below the saturation field of the Heisenberg AF/AF triangular bilayer in the asymptotic limit of zero temperature due to an extra energy penalty. Owing to the symmetry, the polarized triplet states on neighboring dimers also repel each other above the first critical field connected with a breakdown of the singlet-dimer ground state (\ref{sgs}). Hence, it follows that the magnetic behavior of the Heisenberg AF/AF triangular bilayer can be reasonably well approximated at low enough temperatures by a hard-hexagon model on a triangular lattice, which is retrieved from the effective lattice-gas model given by Eq. (\ref{hamlg}) in the limit of infinitely large repulsion $J_1 \to \infty$. The partition function of the spin-$\frac{1}{2}$ Heisenberg triangular AF/AF bilayer at low temperatures can be thus obtained from the grand-canonical partition function of a hard-hexagon model on a triangular lattice:
\begin{eqnarray}
\label{gp}
{\cal Z} (\beta, J_1, J_2) = \exp(-\beta E_{\rm FM}) \Xi_{hh} (\mu),
\end{eqnarray}
where $\mu = J_2 + 6J_1 -h$ ($\mu = h - J_2$) is a chemical potential of the hexagon particles obeying hard-core potential below the third (above the first) critical field 
$h_{c3} = J_2 + 6J_1$ ($h_{c1} = J_2$). The hard-hexagon model on a triangular lattice has been exactly solved due to Baxter \cite{baxt80,baxt82} and the exact result for critical fugacity (activity) $z_{c} = \frac{1}{2} (11 + 5 \sqrt{5})$ affords the following critical conditions for the spin-$\frac{1}{2}$ Heisenberg AF/AF triangular bilayer:
\begin{eqnarray}
T_{c} \!\!\!&=&\!\!\! \frac{h - h_{c1}}{k_{\rm B} \ln z_{c}} \qquad \mbox{for} \quad h \gtrsim  h_{c1} = J_2, \nonumber \\
T_{c} \!\!\!&=&\!\!\! \frac{h_{c3} - h}{k_{\rm B} \ln z_{c}} \qquad \mbox{for} \quad h \lesssim h_{c3} = J_2 + 6J_1, 
\label{tc}
\end{eqnarray}
which determine low-temperature asymptotes of the critical temperature in a close vicinity of the first and third critical fields. 
 
\subsection{Exact diagonalization} 

To corroborate reliability of the developed approach for a description of the low-temperature magnetization process and thermodynamics we have performed a full exact diagonalization (ED) of the spin-$\frac{1}{2}$ Heisenberg triangular bilayer with the linear size $L=3$ and the total number of spins $2N=18$ under the periodic boundary conditions by adapting the subroutines from the ALPS project.\cite{alps11} The ED data for the spin-$\frac{1}{2}$ Heisenberg triangular bilayer with $L=3$ will be confronted with exact results for the effective triangular Ising model with the same linear size $L$ (see Appendix \ref{appb}). The full ED data will bring insight into a range of applicability of the effective Ising model and moreover, they provide useful benchmark for the numerical data obtained from the MC simulations of the effective Ising model on a triangular lattice of much larger linear size. It will be demonstrated hereafter that the ED data of the Heisenberg triangular bilayer with a rather limited size $L=3$ fit surprisingly well results obtained from MC simulations of the effective triangular Ising model for much larger system size (typically $L=180$). 

\section{Results and discussion}
\label{sec:result}

In this section, we will perform a comprehensive analysis of the most interesting results for the low-temperature magnetization process and thermodynamics of the spin-$\frac{1}{2}$ Heisenberg triangular bilayer by considering the antiferromagnetic intradimer interaction ($J_2>0$) and either ferromagnetic ($J_1<0$) or antiferromagnetic ($J_1>0$) interdimer interaction.  

\subsection{FM/AF bilayer ($J_1<0$, $J_2>0$)}  

\begin{figure}
\begin{center}
\includegraphics[width=0.5\textwidth]{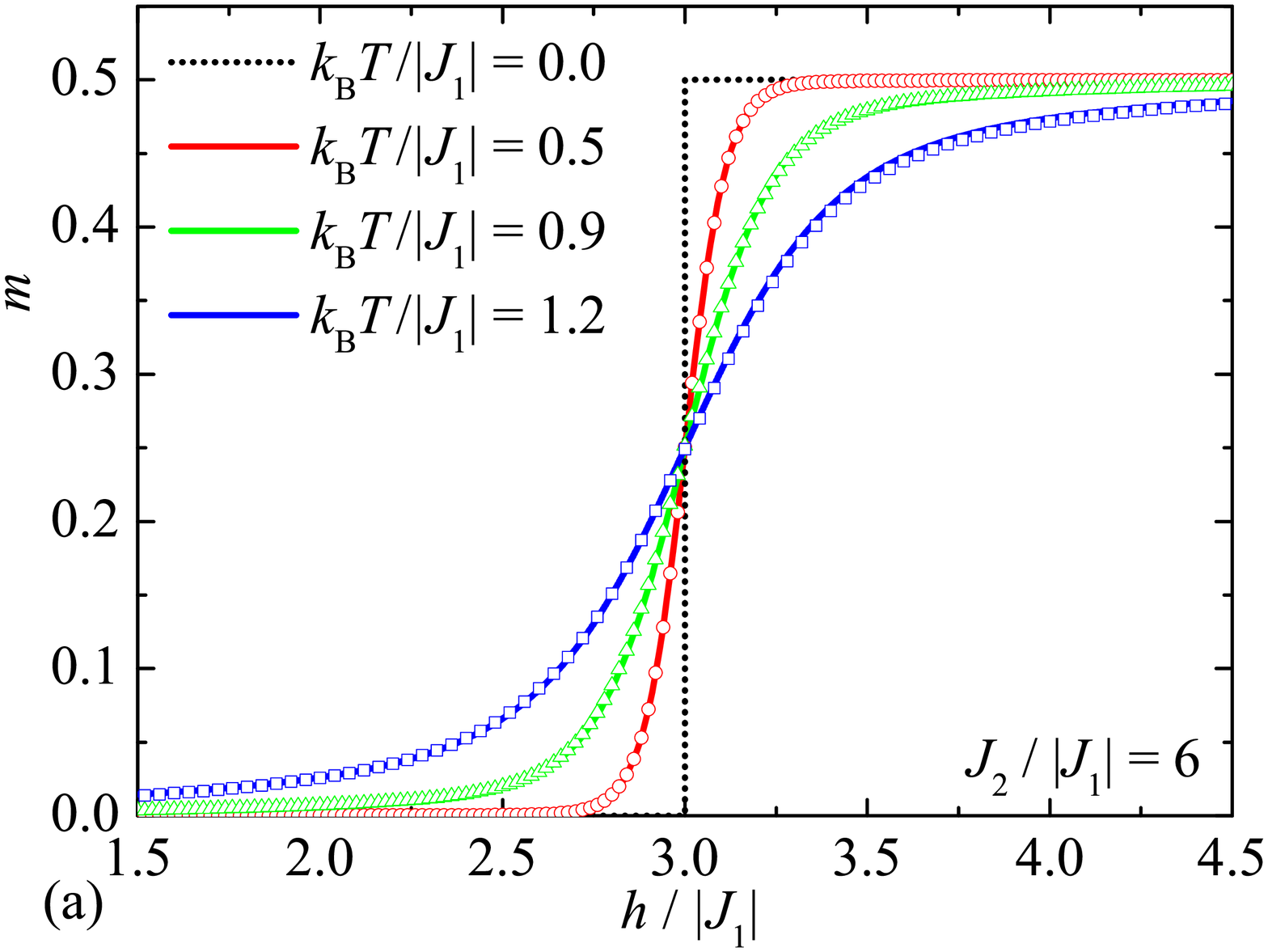}
\includegraphics[width=0.5\textwidth]{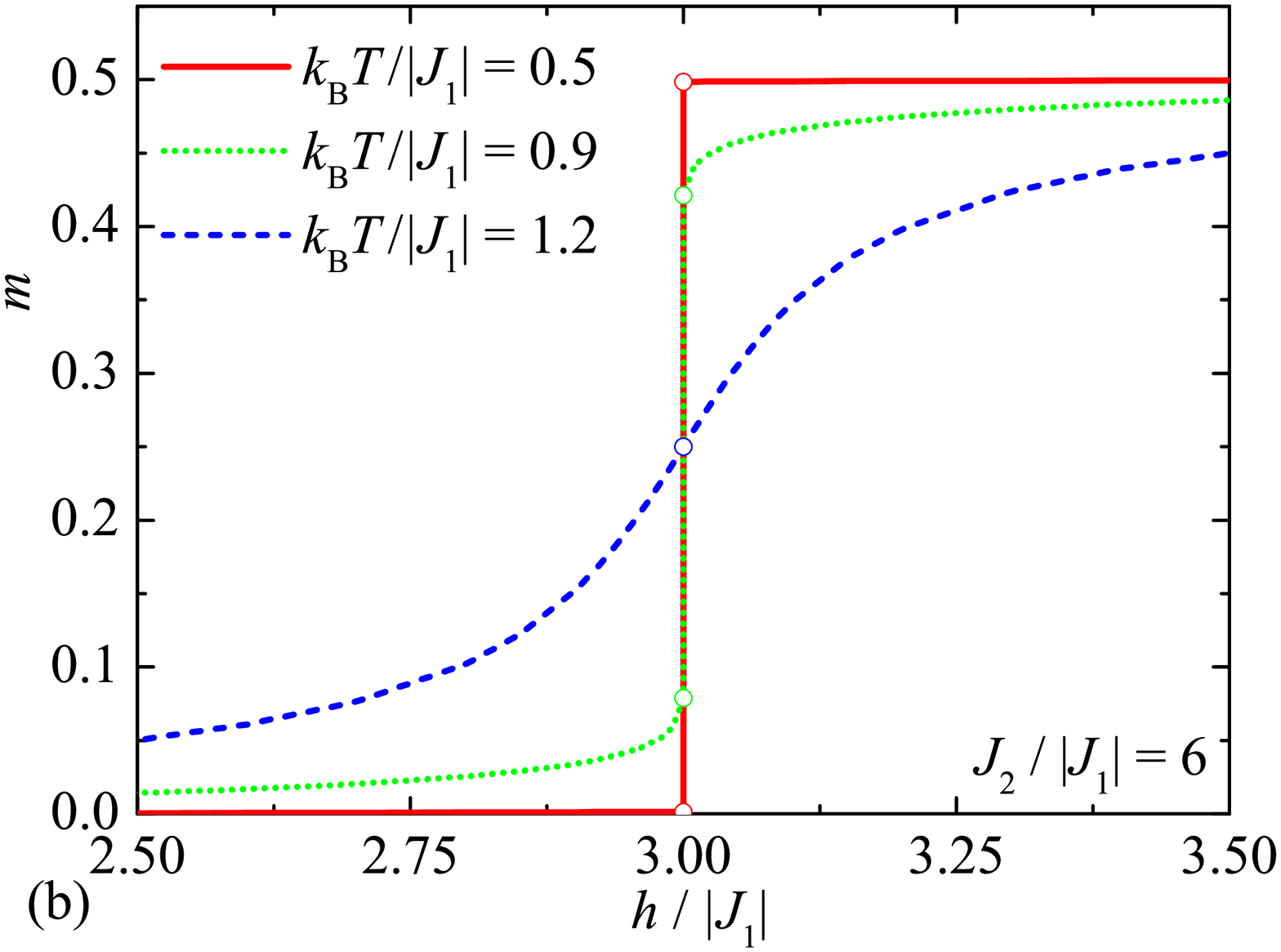}
\end{center}
\vspace{-0.6cm}
\caption{The isothermal magnetization curves of the spin-$\frac{1}{2}$ Heisenberg triangular bilayer with the linear size $L$ for the particular case $J_2/|J_1| = 6$ 
and a few different temperatures. (a) Full ED data for the Heisenberg bilayer with $L=3$ (open symbols) are compared with the exact results for the effective Ising model (solid lines); (b) MC simulations for the effective Ising model with $L=180$. Open symbols display the magnetization at a critical field as obtained from the exact result (\ref{magpt}) of the corresponding Ising model at zero effective field.}
\label{mf}
\end{figure}

At first let us compare exact results for the effective $3\times 3$ triangular Ising ferromagnet with the full ED data for the spin-$\frac{1}{2}$ Heisenberg FM/AF triangular bilayer with $L=3$ (i.e. $2\times3 \times 3 = 18$ spins). The isothermal magnetization curves of both these models are depicted in Fig. \ref{mf}(a) for the interaction ratio $J_2/|J_1|=6$ and a few different temperatures. It is worthwhile to recall that a validity of the localized many-magnon approach is restricted by the condition $J_2/|J_1|>3$ so that the selected value of the interaction ratio $J_2/|J_1|=6$ falls deep inside of this parameter space. The zero-temperature magnetization curve of the spin-$\frac{1}{2}$ Heisenberg FM/AF triangular bilayer with $L=3$ exhibits zero magnetization plateau, which terminates just at the critical field $h_c/|J_1|= J_2/|J_1|-3$ where the magnetization jumps to its saturation value. It should be emphasized that a true magnetization jump does not appear at any finite temperature, because rising temperature generally causes a gradual smoothing of the magnetization curve. It can be seen from Fig. \ref{mf}(a) that the full ED data for the spin-$\frac{1}{2}$ Heisenberg triangular bilayer with $L=3$ are in a perfect agreement with exact results for the effective $3\times 3$ triangular Ising ferromagnet up to moderate temperatures $k_{\rm{B}}T/|J_1|\lesssim 1.2$.

With this background, it is quite plausible to suspect that the magnetization curve of the spin-$\frac{1}{2}$ Heisenberg FM/AF triangular bilayer with much larger system size can be reasonably well approximated at low enough temperatures by the effective triangular Ising ferromagnet. For this purpose, the isothermal magnetization curves as obtained from MC simulations of the ferromagnetic Ising model on a triangular lattice with the linear size $L=180$ are depicted in Fig. \ref{mf}(b) for three different temperatures. The magnetization curve of the spin-$\frac{1}{2}$ Heisenberg FM/AF triangular bilayer still exhibits zero magnetization plateau, but a discontinuous magnetization jump persists at low enough temperatures. It actually turns out that the size of discontinuous magnetization jump is just gradually suppressed upon increasing of temperature until a continuous field-driven phase transition from the Ising universality class is reached at the critical temperature $k_{\rm{B}}T_c/|J_1|=1/\ln 3 \approx 0.91$. Above the critical temperature the magnetization varies continuously upon strengthening of the magnetic field without any type of singularity. It should be pointed out that the numerical results obtained from MC simulations of the effective triangular Ising ferromagnet with the linear size $L = 180$ are in an excellent coincidence with the exact analytical result (\ref{magpt}) for the magnetization available in the thermodynamic limit $L \to \infty$ at the critical field (i.e. zero effective field).

\begin{figure}
\begin{center}
\includegraphics[width=0.5\textwidth]{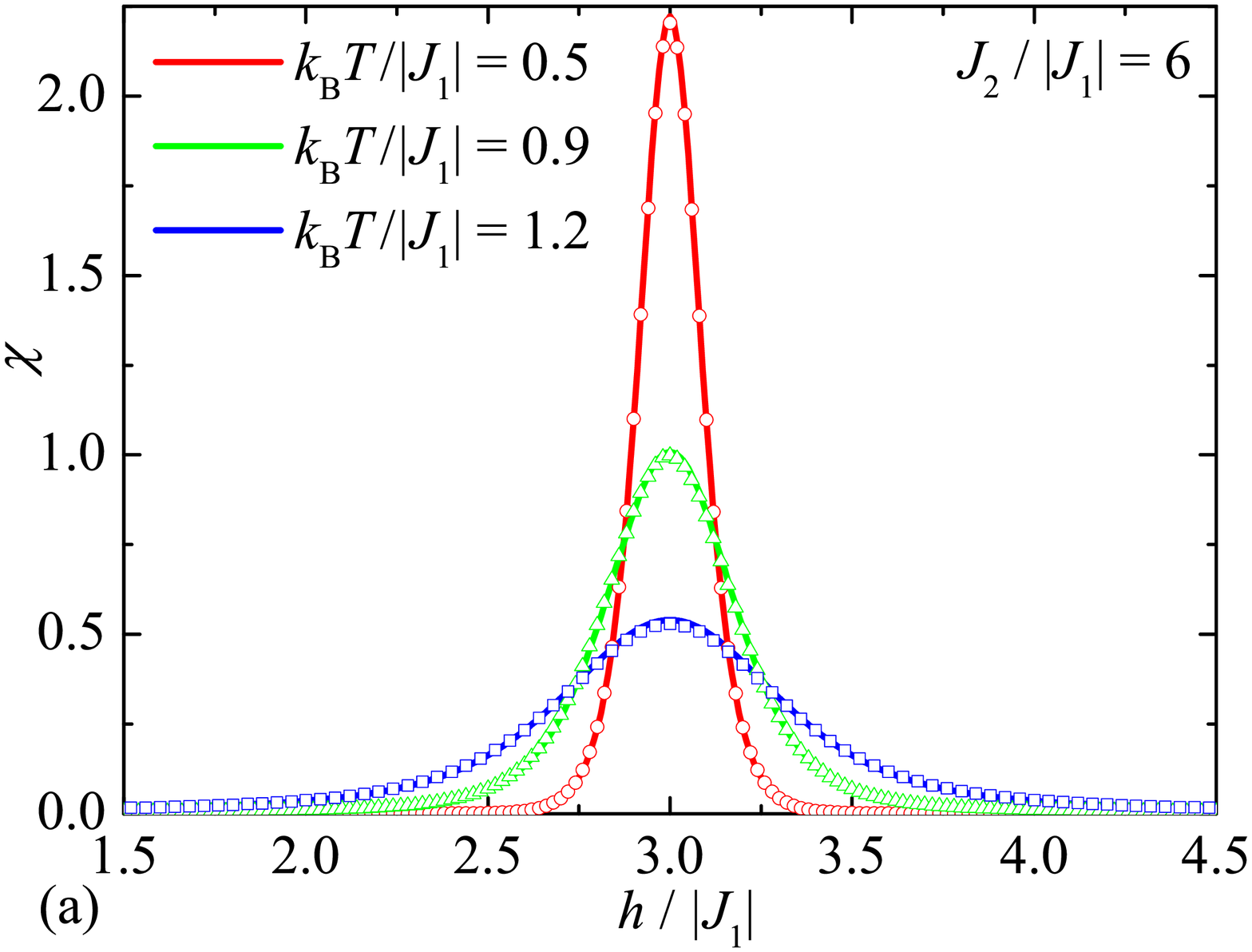}
\includegraphics[width=0.5\textwidth]{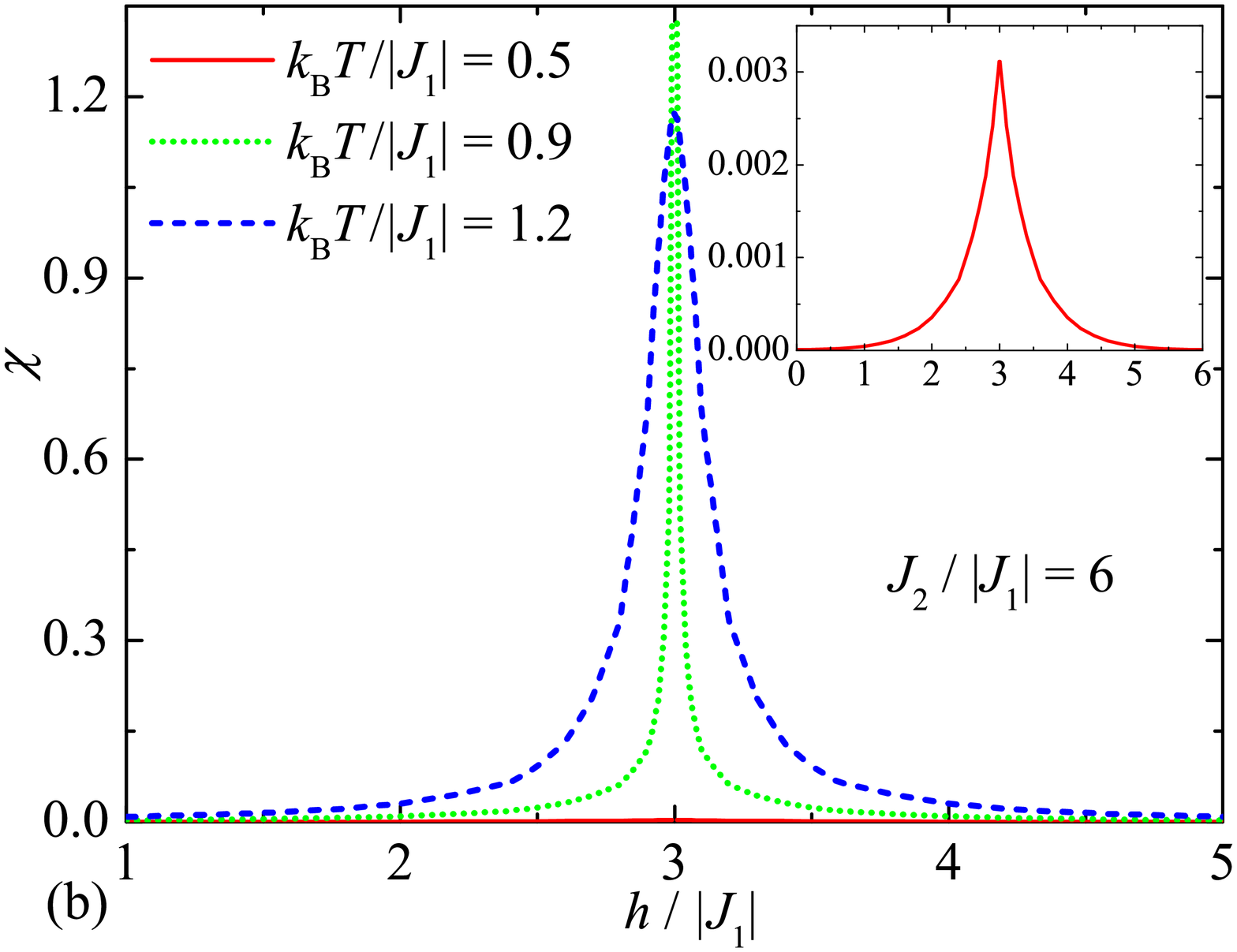}
\end{center}
\vspace{-0.6cm}
\caption{The isothermal field dependence of the susceptibility of the spin-$\frac{1}{2}$ Heisenberg triangular bilayer with the linear size $L$ for the particular case $J_2/|J_1| = 6$ and a few different temperatures. (a) Full ED data for the Heisenberg bilayer with $L=3$ (open symbols) are compared with the exact results for the effective Ising model (solid lines); (b) MC simulations for the effective Ising model with $L=180$. The insert shows the susceptibility at the lowest temperature $k_{\rm{B}}T/|J_1| = 0.5$ in an enhanced scale.}
\label{sf}
\end{figure}

The ED data of the isothermal susceptibility of the spin-$\frac{1}{2}$ Heisenberg FM/AF triangular bilayer with $L=3$ are plotted in Fig. \ref{sf}(a)  against the magnetic field along with exact results for the corresponding effective triangular Ising ferromagnet. As one can see, the susceptibility of the spin-$\frac{1}{2}$ Heisenberg FM/AF triangular bilayer with $L=3$ exhibits at the critical field $h_c/|J_1|= J_2/|J_1|-3$ ($h_{\rm eff}=0$) a round maximum, which becomes higher and sharper upon lowering temperature. The susceptibility data obtained from the effective triangular Ising ferromagnet with $L=3$ coincide with the relevant ED data up to moderate temperatures $k_{\rm{B}}T/|J_1|\lesssim 1.2$. MC simulations of the effective triangular Ising ferromagnet with $L=180$ shown in Fig. \ref{sf}(b) thus bring insight into the susceptibility of the spin-$\frac{1}{2}$ Heisenberg FM/AF triangular bilayer with much larger system size. The isothermal susceptibility of the spin-$\frac{1}{2}$ Heisenberg FM/AF triangular bilayer accordingly displays at sufficiently low temperatures a sharp cusp with discontinuous derivative, which becomes higher and narrower upon increasing temperature until a power-law divergence from the Ising universality class is reached at the critical temperature. Above the critical temperature the susceptibility displays a smooth temperature dependence with a round maximum without any singularity.  

\begin{figure}
\begin{center}
\includegraphics[width=0.5\textwidth]{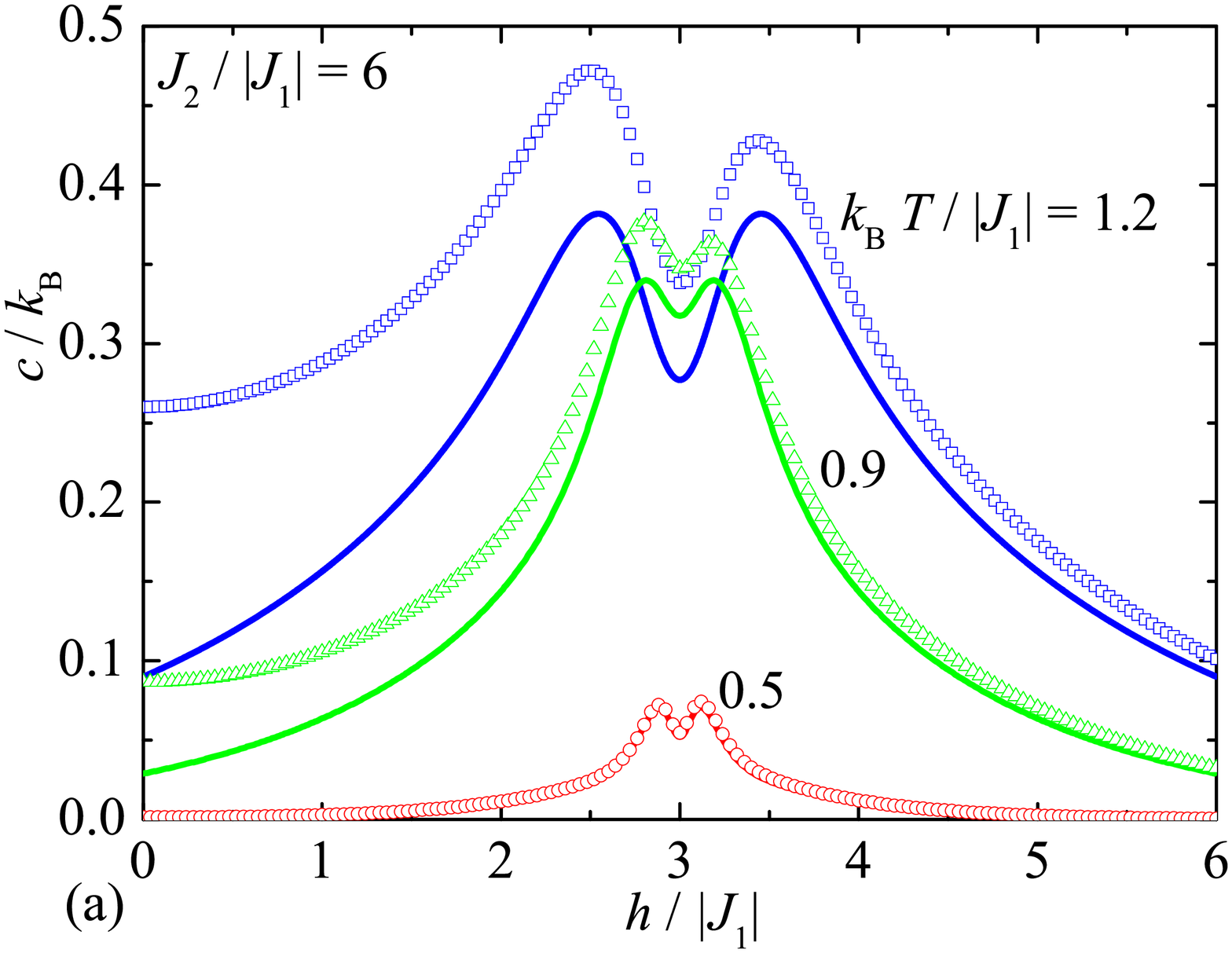}
\includegraphics[width=0.5\textwidth]{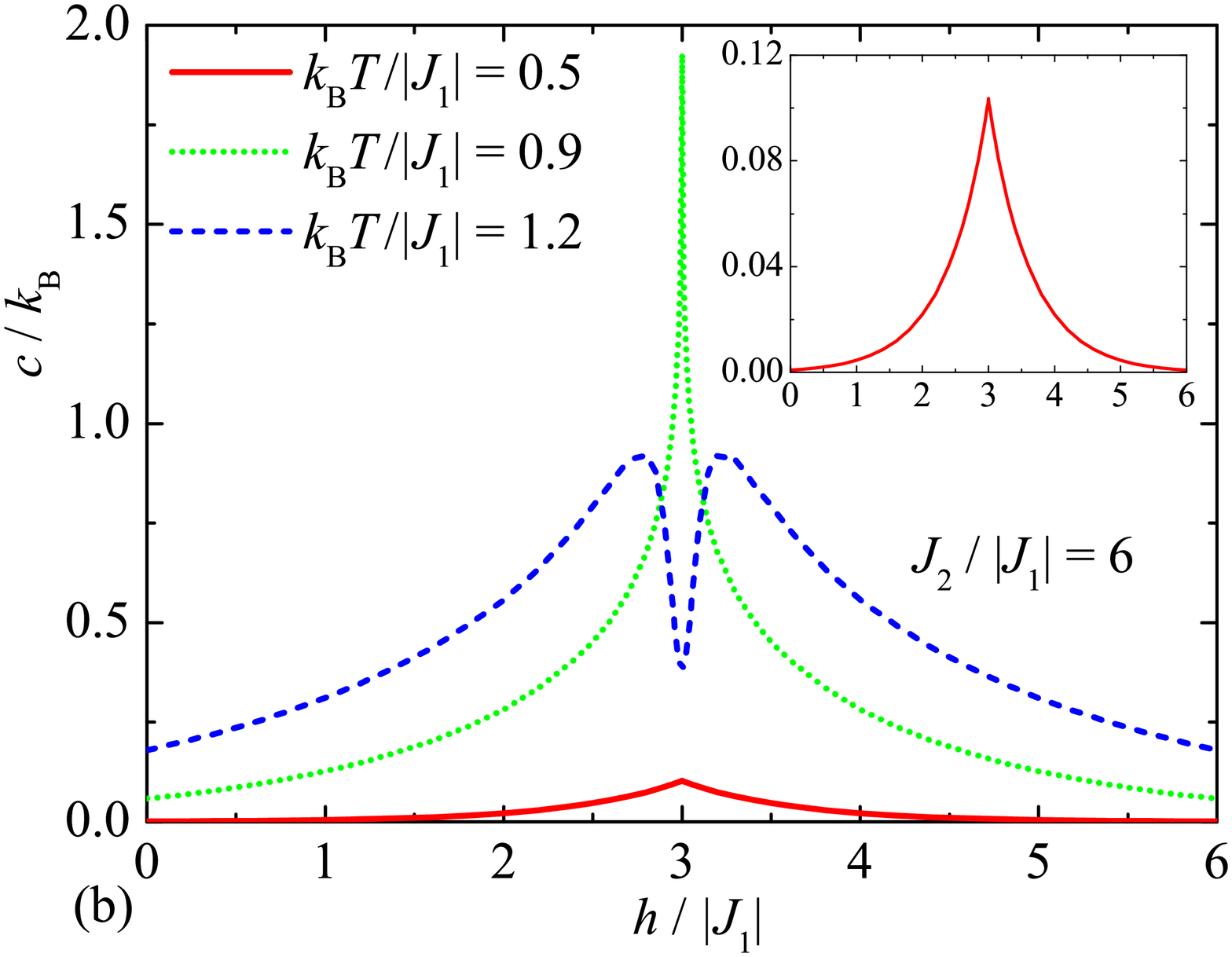}
\end{center}
\vspace{-0.6cm}
\caption{The isothermal field dependence of the specific heat of the spin-$\frac{1}{2}$ Heisenberg triangular bilayer with the linear size $L$ for the particular case $J_2/|J_1| = 6$ and a few different temperatures. (a) Full ED data for the Heisenberg bilayer with $L=3$ (open symbols) are compared with the exact results for the effective Ising model (solid lines); (b) MC simulations for the effective Ising model with $L=180$. The insert shows the specific heat at the lowest temperature $k_{\rm{B}}T/|J_1| = 0.5$ in an enhanced scale.}
\label{fh}
\end{figure}

Last but not least, ED data for the specific heat of the spin-$\frac{1}{2}$ Heisenberg FM/AF triangular bilayer with $L=3$ are depicted in Fig. \ref{fh}(a) as a function of the magnetic field together with exact results derived from the effective $3 \times 3$ triangular Ising ferromagnet. Although the results stemming from both these models display qualitatively the same temperature dependencies with a double-peak structure of the specific heat around a critical field, the reliable quantitative match between the data is found just at lower temperatures $k_{\rm{B}}T/|J_1| \lesssim 0.6$. It actually follows from Fig. \ref{fh}(a) that a height of the double peak as well as the zero-field limit of the specific heat as obtained from the effective triangular Ising ferromagnet (solid lines) are slightly underestimated above temperature $k_{\rm{B}}T/|J_1| \gtrsim 0.6$ in comparison with full ED data of the spin-$\frac{1}{2}$ Heisenberg triangular bilayer with $L=3$ even though a position of double peaks is still adequate. A physical origin of two peaks emergent in a vicinity of the critical field lies in vigorous thermal excitations of vertical dimers from a singlet ground state towards a low-lying polarized triplet state (a peak at $h < h_c$) or vice versa (a peak at $h > h_c$).

Temperature variations of the specific heat as obtained from MC simulations of the effective Ising model on a triangular lattice with linear size $L=180$ are plotted in Fig. \ref{fh}(b) in order to shed light on the respective behavior of the spin-$\frac{1}{2}$ Heisenberg triangular bilayer of a much larger system size. The specific heat of the frustrated spin-$\frac{1}{2}$ Heisenberg FM/AF triangular bilayer displays at low temperatures a finite cusp, which increases in height upon increasing temperature until a logarithmic divergence from the Ising universality class is reached at the critical temperature $k_{\rm{B}}T_c/|J_1|=1/\ln 3 \approx 0.91$. The specific heat thus displays at low enough temperatures essential differences in the relevant magnetic-field dependencies in comparison with the relevant behavior of small-size systems [c.f. Fig. \ref{fh}(a) and (b)], which can be attributed to a cooperative nature of the spontaneous long-range order of the singlet and polarized triplet states of the vertical dimers that is of course elusive for small-size systems. Above the critical temperature the magnetic-field dependencies of the specific heat are reminiscent of the ones of small-system sizes with two round maxima emergent close to a critical field, because the spontaneous long-range order of the singlet and polarized triplet states is absent and there are just short-range correlations in their abundance that are quite typical also for small-size systems.  

\subsection{AF/AF bilayer ($J_1>0$, $J_2>0$)}

\begin{figure}
\begin{center}
\includegraphics[width=0.5\textwidth]{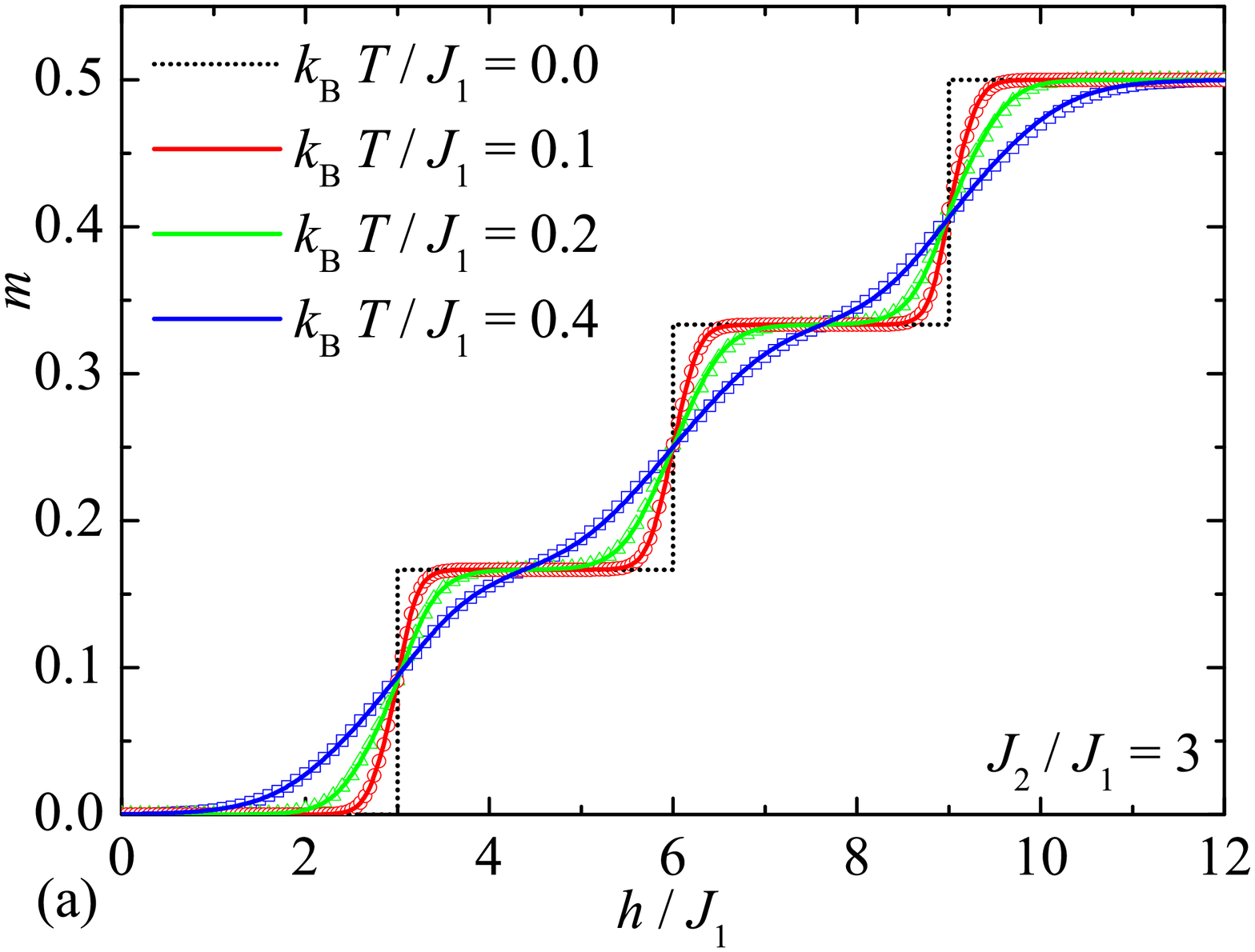}
\includegraphics[width=0.5\textwidth]{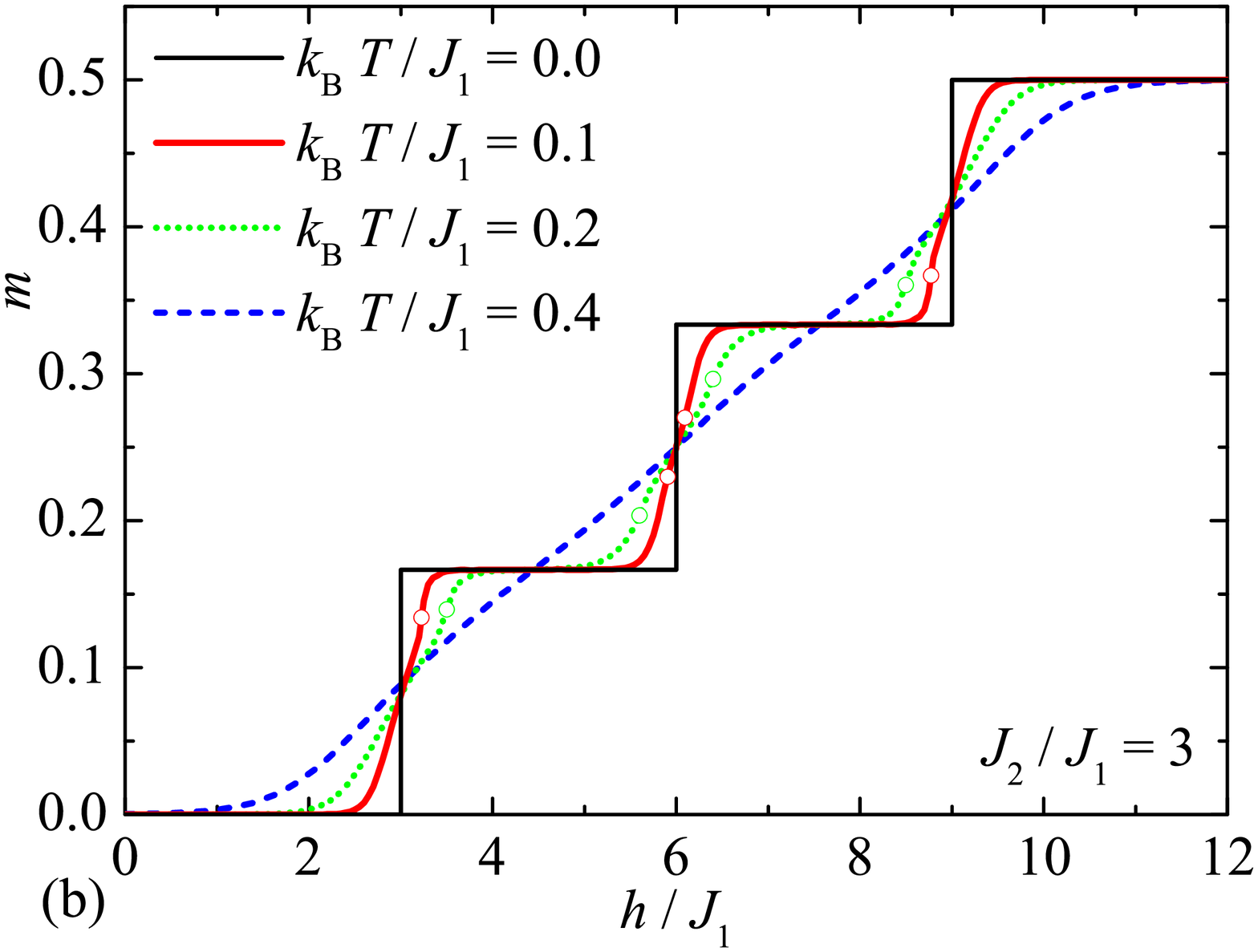}
\end{center}
\vspace{-0.6cm}
\caption{The isothermal magnetization curves of the spin-$\frac{1}{2}$ Heisenberg triangular bilayer with the linear size $L$ for the particular case $J_2/J_1 = 3$ and a few different temperatures. (a) Full ED data for the Heisenberg bilayer with $L=3$ (open symbols) are compared with the exact results for the effective Ising model (solid lines); (b) MC simulations of the effective Ising model with $L=180$. Open circles denote critical points.}
\label{maf}
\end{figure}

In the following part we will investigate a magnetic behavior of the frustrated spin-$\frac{1}{2}$ Heisenberg triangular bilayer by considering the antiferromagnetic interdimer ($J_1>0$) and intradimer ($J_2>0$) interactions. ED data for the isothermal magnetization curves of the spin-$\frac{1}{2}$ Heisenberg AF/AF triangular bilayer with the linear size $L = 3$ are confronted in Fig. \ref{maf}(a) with exact results for the effective triangular Ising antiferromagnet. According to this plot, the stepwise magnetization curve with intermediate plateaux at zero, one-third and two-thirds of the saturation magnetization and the respective magnetization jumps observable strictly at zero temperature are gradually smeared out upon increasing of temperature. Besides, the results presented in Fig. \ref{maf}(a) serve in evidence that a description based on the effective triangular Ising antiferromagnet is faithful up to moderate temperatures $k_{\rm{B}}T/|J_1| \lesssim 0.5$. 

Bearing this in mind, the results derived from MC simulations of the effective triangular Ising antiferromagnet with much larger linear size $L=180$ should provide a reliable estimate of the isothermal magnetization curves of the spin-$\frac{1}{2}$ Heisenberg triangular bilayer in this temperature range [see Fig. \ref{maf}(b)]. The magnetization curve of the frustrated spin-$\frac{1}{2}$ Heisenberg AF/AF triangular bilayer accordingly displays intermediate magnetization plateaux at zero, one-third and two-thirds of the saturation magnetization, whereas the latter two magnetization plateaux are conformable with two aforedescribed period-three ground states with a regular alternation of 'singlet-singlet-triplet' and 'singlet-triplet-triplet' dimer states, respectively. Most strikingly, the magnetization curve of the frustrated spin-$\frac{1}{2}$ Heisenberg AF/AF triangular bilayer involves at sufficiently low temperatures four singular points shown in Fig. \ref{maf}(b) as open circles, which bear evidence of  continuous field-driven phase transitions between the individual ground states. 

\begin{figure}
\begin{center}
\includegraphics[width=0.5\textwidth]{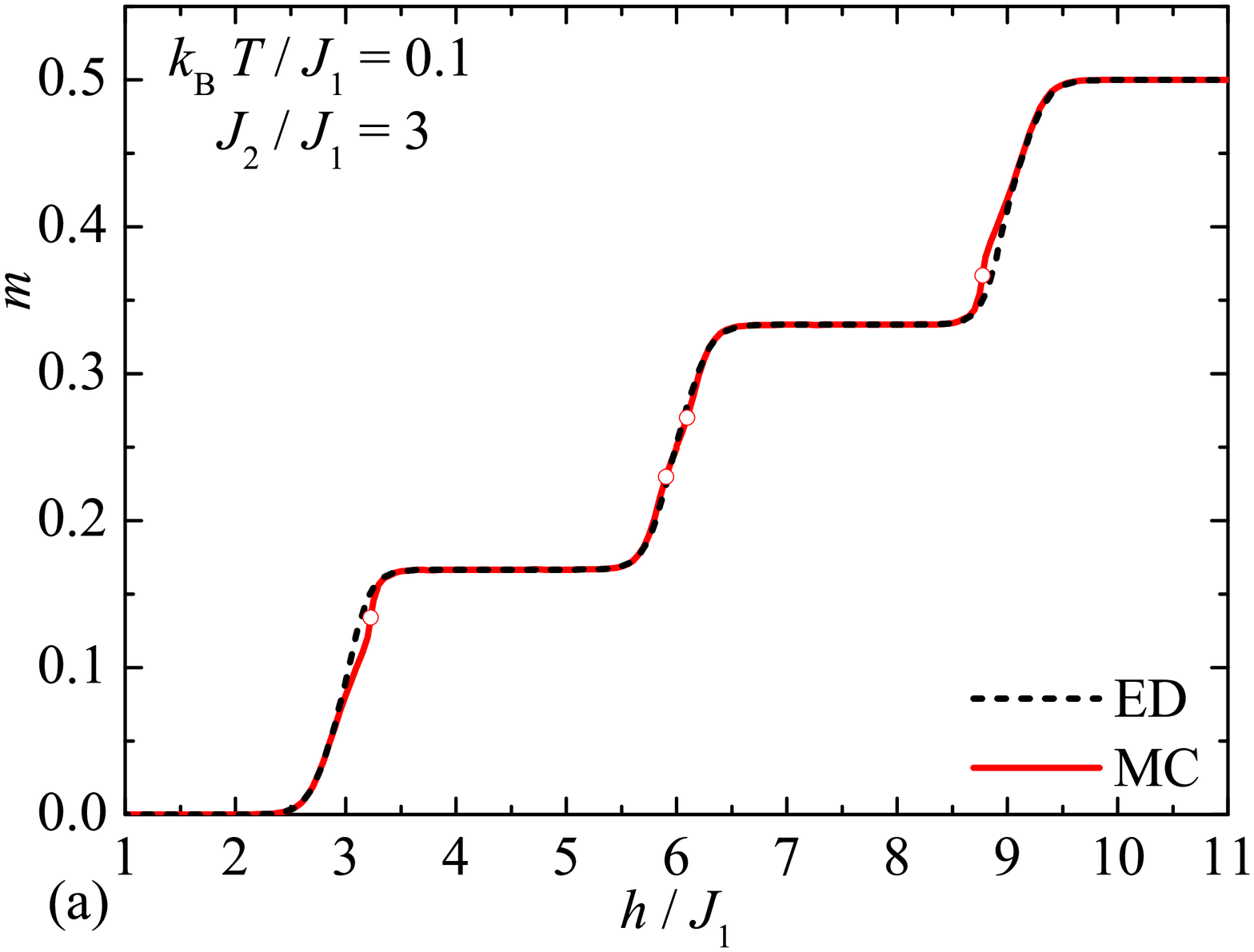}
\includegraphics[width=0.5\textwidth]{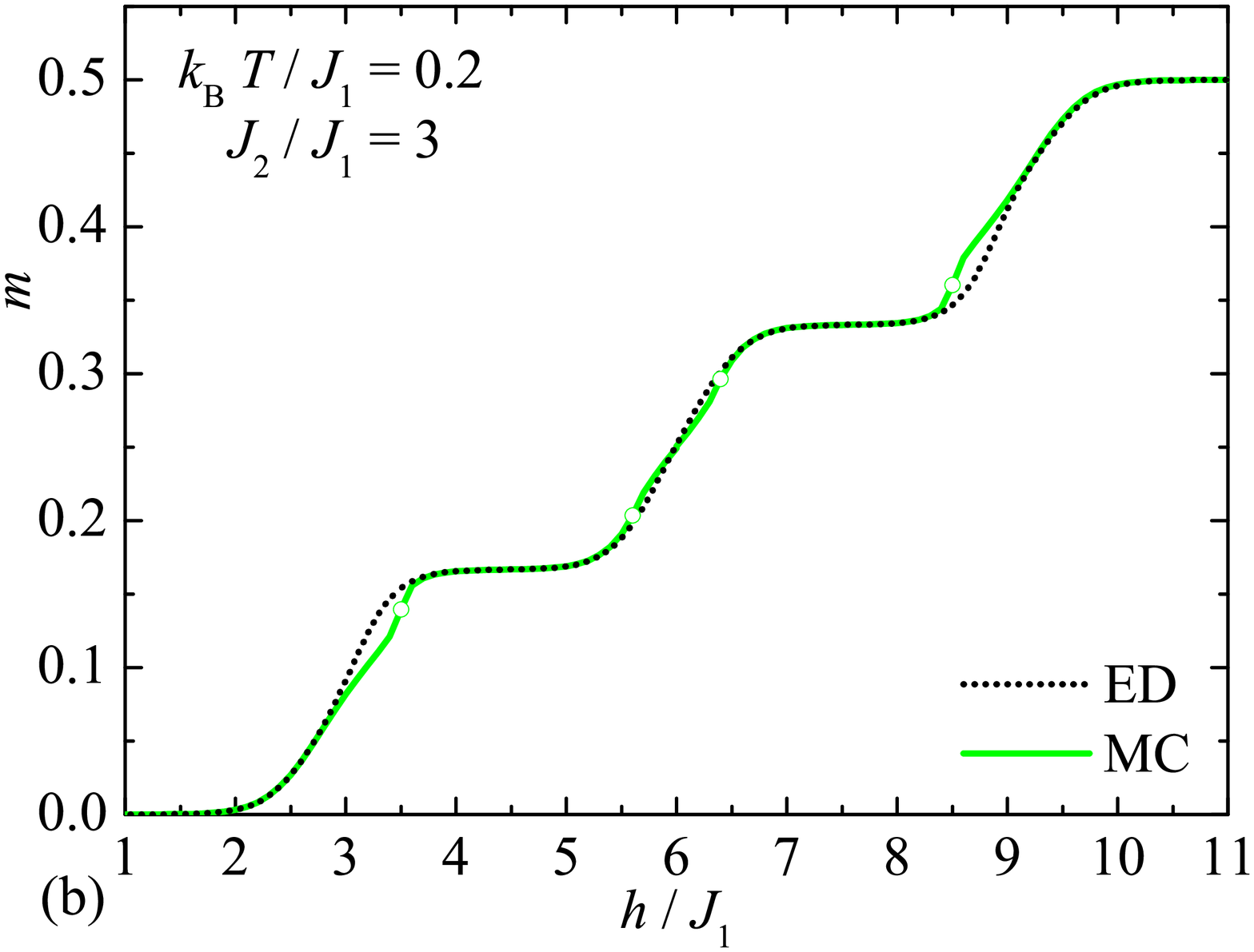}
\end{center}
\vspace{-0.6cm}
\caption{A comparison of the isothermal magnetization curves of the spin-$\frac{1}{2}$ Heisenberg triangular bilayer with $J_2/J_1=3$ as obtained from full ED calculations for $L=3$ and  MC simulations of the effective Ising model for $L=180$ at two different temperatures: (a) $k_{\rm B} T/J_1 = 0.1$; (b) $k_{\rm B} T/J_1 = 0.2$.}
\label{mafc}
\end{figure}

It might be quite helpful to examine a difference between the magnetization curve of the spin-$\frac{1}{2}$ Heisenberg AF/AF triangular bilayer at two markedly different system sizes. The full ED data for the magnetization curves of the spin-$\frac{1}{2}$ Heisenberg triangular bilayer with the linear size $L = 3$ are consequently compared in Fig. \ref{mafc} with MC simulations of the effective triangular Ising antiferromagnet of much larger linear size $L=180$. A sound quantitative agreement notwithstanding of a considerable difference in a system size is rather surprising. The only substantial difference between the magnetization curves eventually appears in a close vicinity of the continuous field-driven phase transitions, which are of course missing in the relevant magnetization curves for small system sizes such as $L=3$ displaying at any nonzero temperature only a crossover phenomenon instead of the actual field-induced phase transitions.   

\begin{figure}
\begin{center}
\includegraphics[width=0.5\textwidth]{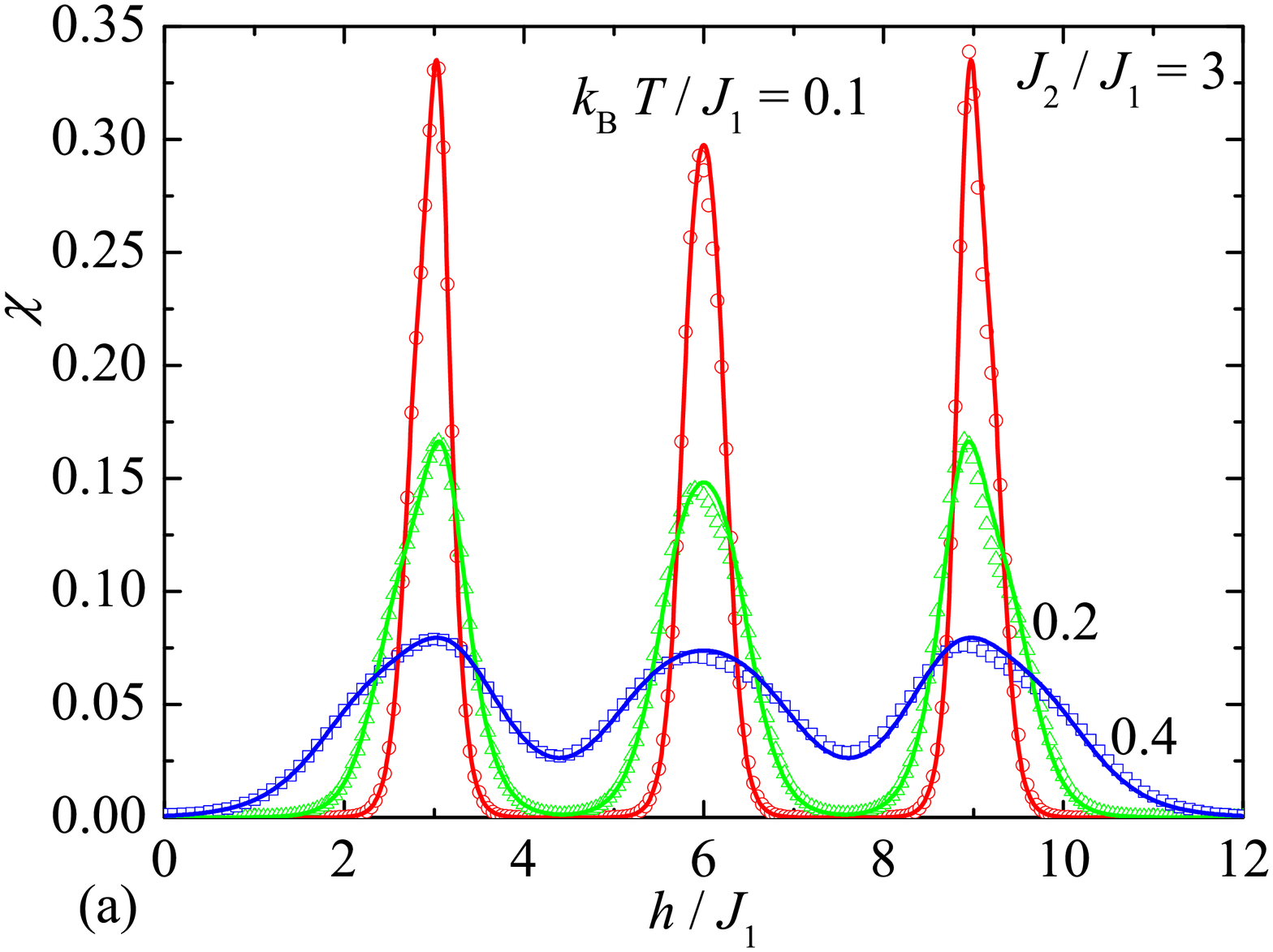}
\includegraphics[width=0.5\textwidth]{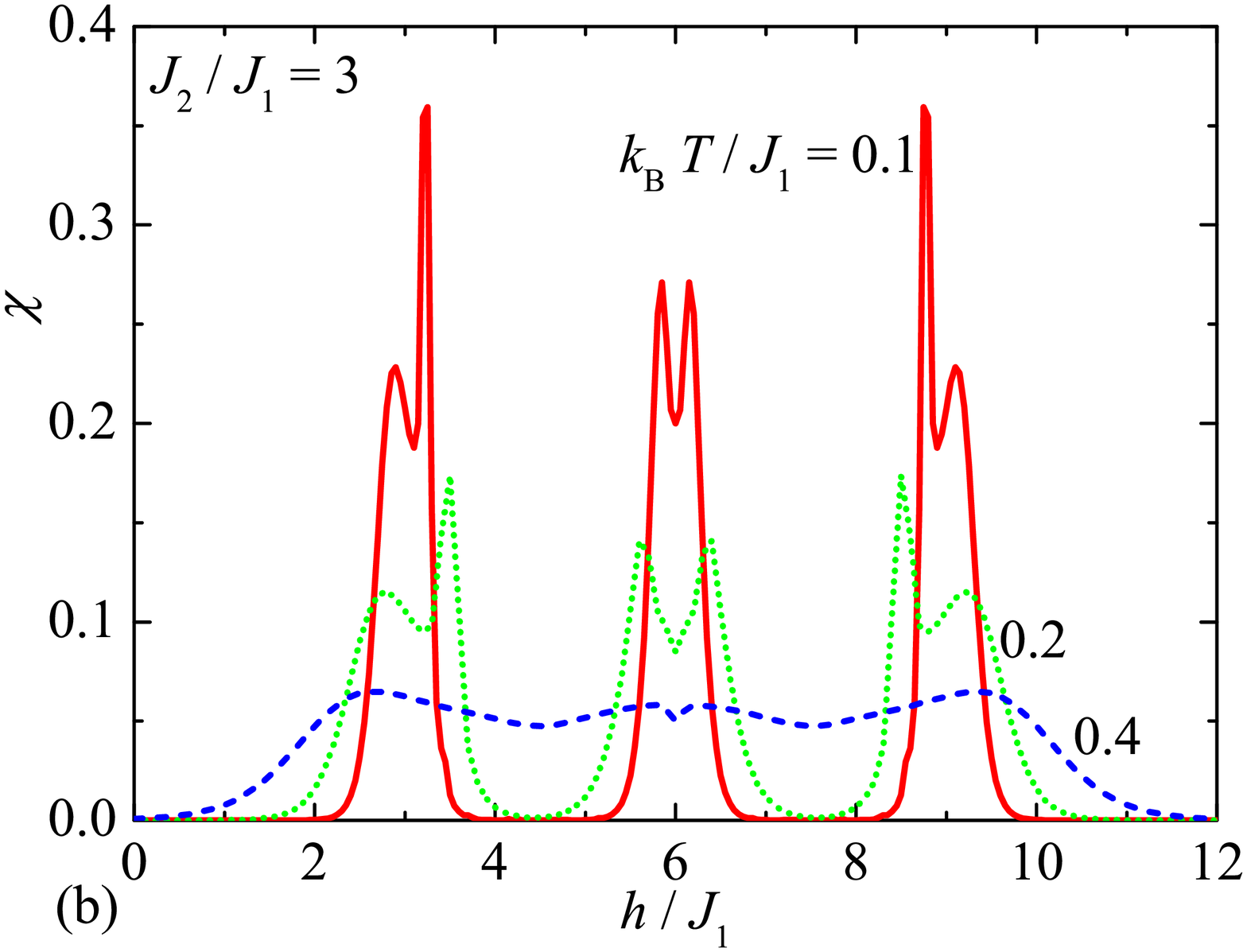}
\end{center}
\vspace{-0.6cm}
\caption{The isothermal field dependence of the susceptibility data of the spin-$\frac{1}{2}$ Heisenberg triangular bilayer for the particular case $J_2/J_1 = 3$ and a few different temperatures. (a) Full ED data for the Heisenberg bilayer with $L=3$ (open symbols) are compared with the exact results for the effective Ising model (solid lines); (b) MC simulations of the effective Ising triangular lattice with $L=180$.}
\label{saf}
\end{figure}

Next, our attention will be paid to magnetic-field variations of the susceptibility of the spin-$\frac{1}{2}$ Heisenberg triangular bilayer with the linear size $L=3$, which were calculated using the full ED method and exact calculations for the effective triangular Ising antiferromagnet, respectively. It can be seen from Fig. \ref{saf}(a) that the results acquired from both these rigorous techniques are in a reasonable accordance up to moderate temperatures $k_{\rm{B}}T/|J_1| \lesssim 0.5$. In fact, a round maximum of the susceptibility allocated at each transition field gradually diminishes upon increasing of temperature, whereas a description based on the effective triangular Ising antiferromagnet correctly reproduces the peak's height as well as position. It could be therefore anticipated that MC simulations of the effective triangular Ising antiferromagnet with the linear size $L=180$ presented in Fig. \ref{saf}(b) afford a proper description of the susceptibility of the spin-$\frac{1}{2}$ Heisenberg triangular bilayer with much larger system size. Obviously, the susceptibility of the larger system size displays at sufficiently low temperatures a markedly different dependence on a magnetic field compared to its small-size counterpart due to a critical behavior accompanying each field-induced phase transition. The susceptibility accordingly diverges at four field-driven phase transitions, whereas one also detects two round maxima located below the first and above the fourth transition field. The latter two round maxima indicate low-lying excitations out of the singlet-dimer and the fully polarized ferromagnetic ground states. Note furthermore that similar findings have been reported on previously also for the fully frustrated spin-$\frac{1}{2}$ Heisenberg square bilayer.\cite{rich06,derz11} 

\begin{figure}
\begin{center}
\includegraphics[width=0.5\textwidth]{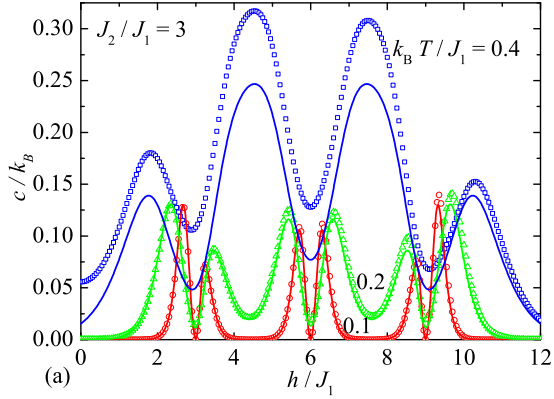}
\includegraphics[width=0.5\textwidth]{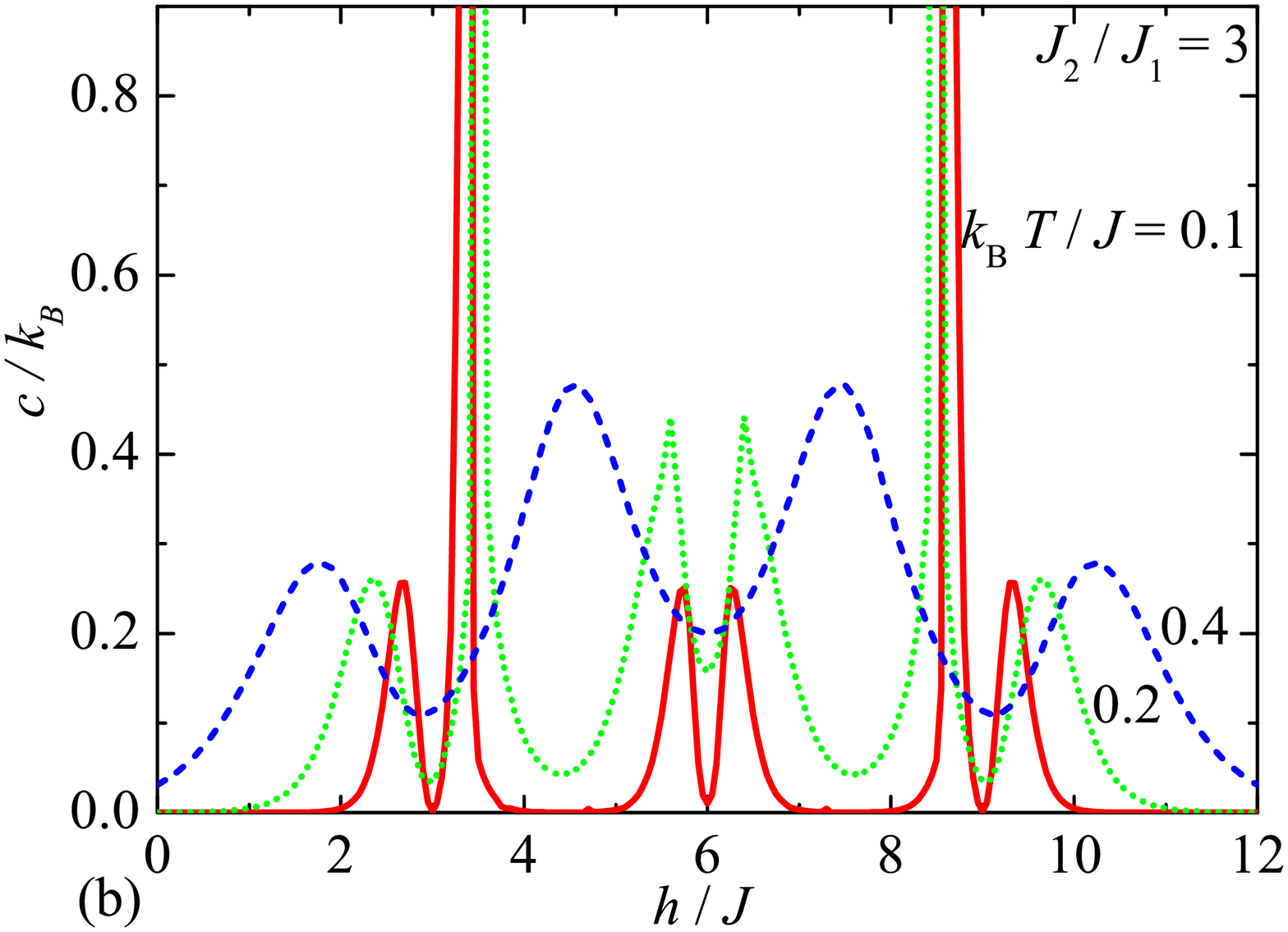}
\end{center}
\vspace{-0.6cm}
\caption{The isothermal field dependence of the specific heat of the spin-$\frac{1}{2}$ Heisenberg triangular bilayer for the particular case $J_2/J_1 = 3$ and a few different temperatures. (a) Full ED data for the Heisenberg bilayer with $L=3$ (open symbols) are compared with the exact results for the effective Ising model (solid lines); (b) MC simulations of the effective Ising triangular lattice with $L=180$.}
\label{haf}
\end{figure}

Last but not least, let us investigate typical magnetic-field dependencies of the specific heat of the spin-$\frac{1}{2}$ Heisenberg triangular bilayer, which are plotted in Fig. \ref{haf}(a) for the finite-size bilayer with the linear size $L=3$. It is quite obvious that the specific heat exhibits a remarkable field dependence with a sequence of three double peaks emerging in a vicinity of the transition fields. Note furthermore that ED data for the Heisenberg triangular bilayer with the linear size $L=3$ are in a feasible quantitative accordance with exact results of the effective triangular Ising antiferromagnet only at lower temperatures $k_{\rm{B}}T/|J_1| \lesssim 0.2$, because the specific heat at higher temperatures is underestimated by the effective triangular Ising model on account of neglected energy levels. 

\begin{figure}
\begin{center}
\includegraphics[width=0.5\textwidth]{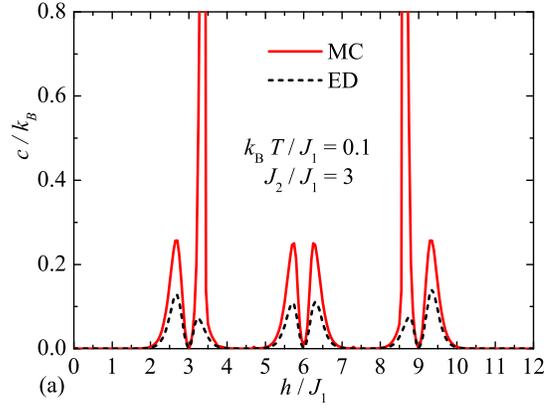}
\includegraphics[width=0.5\textwidth]{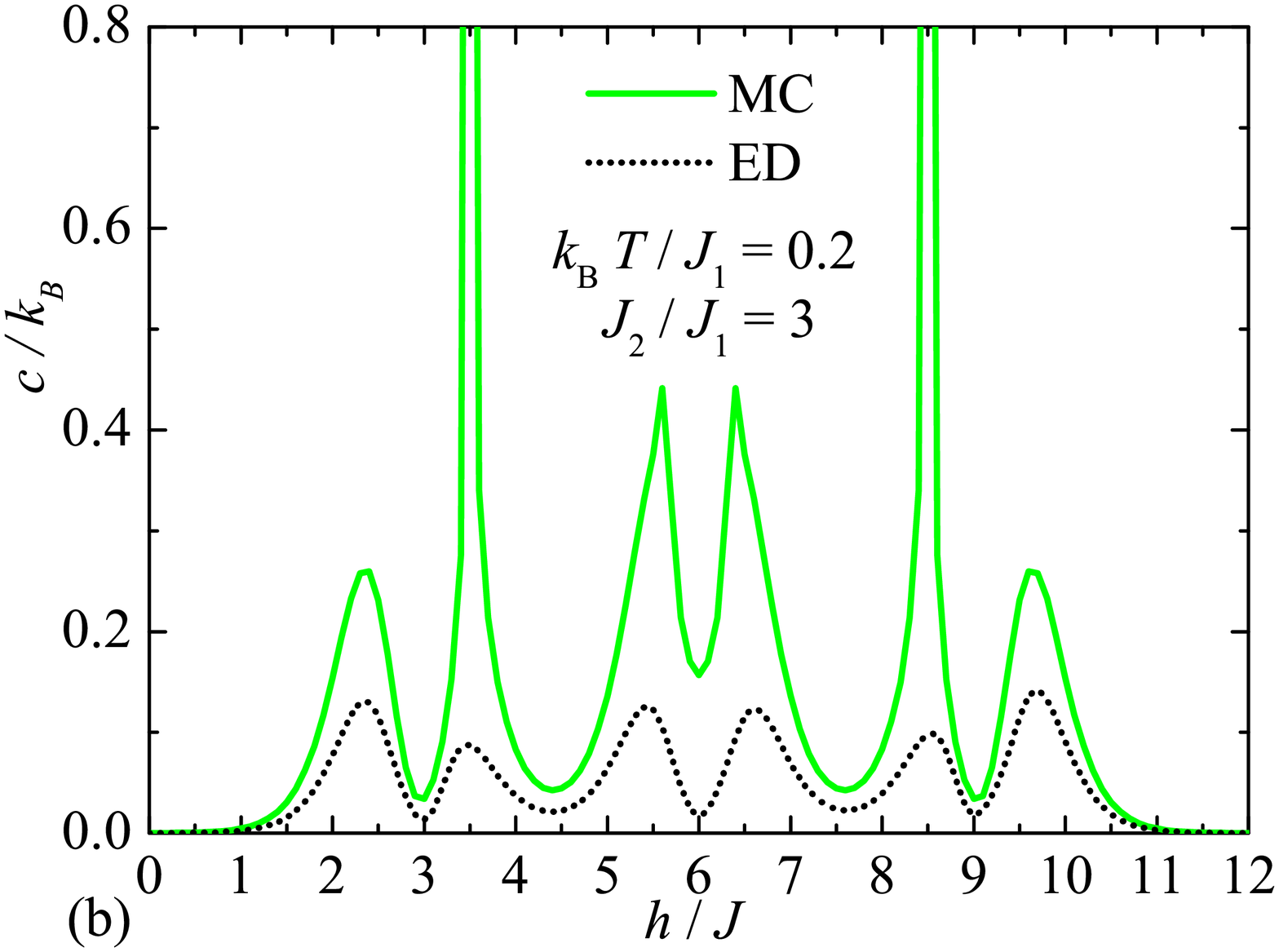}
\end{center}
\vspace{-0.6cm}
\caption{A comparison of the specific heat of the spin-$\frac{1}{2}$ Heisenberg triangular bilayer with $J_2/J_1=3$ as obtained from full ED calculations for $L=3$ and MC simulations of the effective Ising model for $L=180$ at two different temperatures: (a) $k_{\rm B} T/J_1 = 0.1$; (b) $k_{\rm B} T/J_1 = 0.2$.}
\label{hafc}
\end{figure}

The results based on MC simulations of the effective triangular Ising antiferromagnet with the linear size $L=180$ shown in Fig. \ref{haf}(b) should thus provide a reliable estimate of the specific heat of the spin-$\frac{1}{2}$ Heisenberg triangular bilayer at least at low enough temperatures. The magnetic-field dependence of the specific heat of the spin-$\frac{1}{2}$ Heisenberg triangular bilayer consequently exhibits at low enough temperatures four marked divergences from the universality class of three-state Potts model, whereas two additional round maxima can be detected below the first and above the fourth field-driven phase transition. To gain an insight into a cooperative nature of the field-driven phase transitions, the specific heat of the spin-$\frac{1}{2}$ Heisenberg triangular bilayer as obtained from MC simulations of the effective triangular Ising antiferromagnet with the linear size $L=180$ is compared in Fig. \ref{hafc} with the full ED data of the spin-$\frac{1}{2}$ Heisenberg triangular bilayer with the linear size $L=3$. As one can see, the relevant temperature dependencies of the specific heat have apparent similarities as far as the position of emergent maxima is concerned. The main difference thus lies in a height of the specific-heat maxima, which is of course finite for any finite-size system but they rise steadily upon increasing of the system size at the field-driven phase transitions. 

\subsection{Phase diagrams of FM/AF and AF/AF bilayers}

\begin{figure}
\begin{center}
\includegraphics[width=0.5\textwidth]{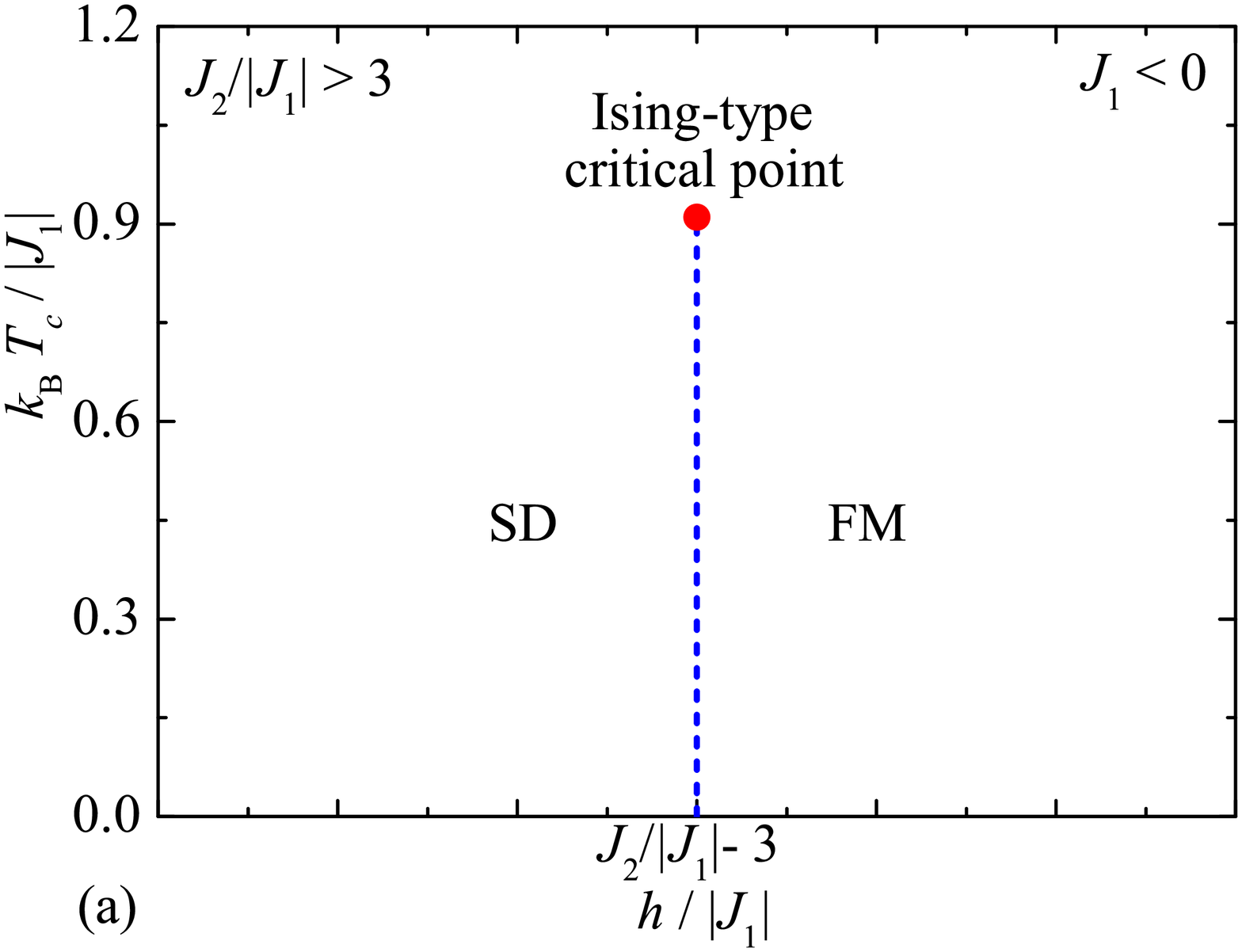}
\includegraphics[width=0.5\textwidth]{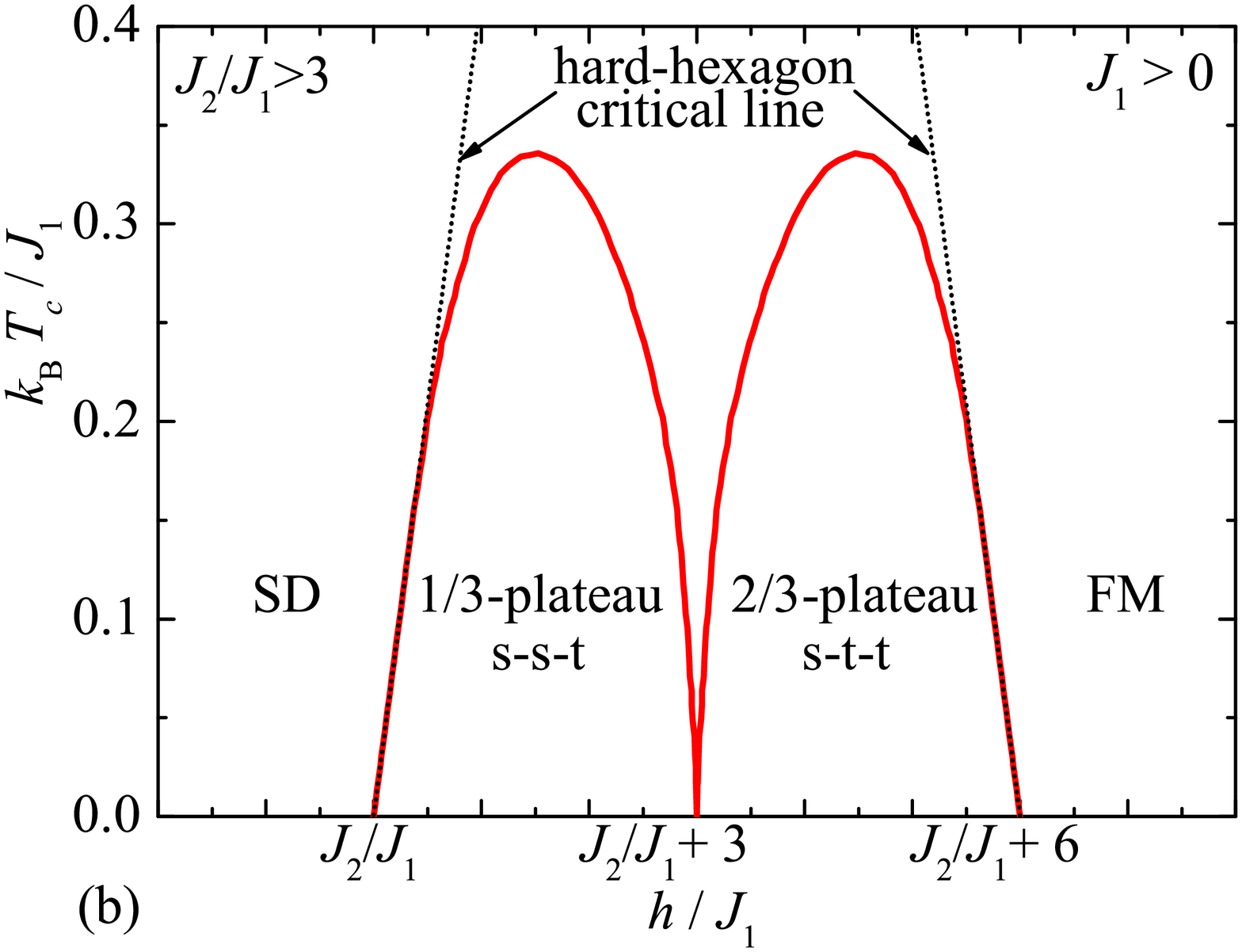}
\end{center}
\vspace{-0.6cm}
\caption{The global phase diagram of the frustrated spin-$\frac{1}{2}$ Heisenberg triangular bilayer in the field-temperature plane as obtained from the exact analytical results for the effective spin-$\frac{1}{2}$ Ising model on a triangular lattice with the ferromagnetic interdimer interaction $J_1<0$ [Fig. \ref{pdfaf}(a)] and the phenomenological scaling adapted from Ref. \onlinecite{kinz81} for the effective spin-$\frac{1}{2}$ Ising model on a triangular lattice with antiferromagnetic interdimer interaction $J_1>0$ [Fig. \ref{pdfaf}(b)]. A broken line in Fig.~\ref{pdfaf}(a) allocates discontinuous field-driven phase transitions terminating at a critical point from the Ising universality class, while solid lines in Fig.~\ref{pdfaf}(b) allocate continuous field-driven phase transitions approaching at sufficiently low temperatures critical boundaries (dotted lines) of a hard-hexagon model on a triangular lattice with the universality class of three-state Potts model. Two domes correspond to intermediate one-third and two-thirds magnetization plateaus with a regular alternation of singlet-singlet-triplet (s-s-t) and singlet-triplet-triplet (s-t-t) dimer states, respectively.}
\label{pdfaf}
\end{figure}

Let us conclude our survey of the most interesting results by constructing global phase diagrams of the frustrated spin-$\frac{1}{2}$ Heisenberg triangular bilayer in the field-temperature plane as obtained from a mapping correspondence (\ref{hamef}), (\ref{mp}) with the effective Ising model on a triangular lattice.  The phase diagram of the spin-$\frac{1}{2}$ Heisenberg FM/AF triangular bilayer can be descended from exact results for the effective triangular Ising ferromagnet,\cite{hout50} while the phase diagram of the spin-$\frac{1}{2}$ Heisenberg AF/AF triangular bilayer has been derived by making use of the results of phenomenological scaling reported in Ref. \onlinecite{kinz81} (MC results reported in the present work are within error bars consistent with this critical line). The global phase diagram of the frustrated spin-$\frac{1}{2}$ Heisenberg FM/AF triangular bilayer with the ferromagnetic interdimer interaction $J_1<0$ depicted in Fig.~\ref{pdfaf}(a) involves a special critical point terminating a vertical line of discontinuous field-driven phase transitions between two nondegenerate phases, which are accompanied with an abrupt magnetization jump emerging at the critical field $h_c/|J_1| = J_2/|J_1| - 3$. The abrupt magnetization jump, which reflects a direct field-driven phase transition from the singlet-dimer phase towards the classical ferromagnetic phase without any intermediate state with a fractional value of the magnetization, bears relation to the ferromagnetic character of the interdimer interaction $J_1<0$ that favors eigenstates with identical states of the dimeric unit cell. It should be stressed, however, that a size of the magnetization jump diminishes upon increasing of temperature until a continuous field-driven phase transition is reached at a special critical point from the Ising universality class with the locus $[h_c/|J_1| ; k_{\rm B} T_c /|J_1|]  = [J_2/|J_1| - 3 ; 1/\ln 3]$. It is worthwhile to recall that the isothermal magnetization curve at higher temperatures (i.e. $T>T_c$) is free from any magnetization discontinuities or singularities. 

The global phase diagram of the frustrated spin-$\frac{1}{2}$ Heisenberg AF/AF triangular bilayer with the antiferromagnetic interdimer interaction $J_1>0$ is much more complex, because it involves apart from the singlet-dimer and ferromagnetic phases two additional quantum phases with a period-three alternation of singlet and polarized triplet states [see Fig.~\ref{pdfaf}(b)]. Consequently, there appear at zero temperature three discontinuous magnetization jumps at the critical fields $h_{c1}/J_1 = J_2/J_1$, $h_{c2}/J_1 = J_2/J_1 + 3$, and $h_{c3}/J_1 = J_2/J_1 + 6$, which are however replaced by four different continuous field-induced phase transitions at finite (nonzero) temperatures. It is noteworthy that the global phase diagram is symmetric with respect to the second critical field $h_{c2}/J_1 = J_2/J_1 + 3$, because this particular value of the magnetic field corresponds to a zero effective field ($h_{\rm eff} =0$) of the effective triangular Ising antiferromagnet. It is quite obvious from Fig.~\ref{pdfaf}(b) that the first and fourth critical fields are approaching at low enough temperatures critical boundaries (\ref{tc}) of a hard-hexagon model on a triangular lattice (dotted lines), which bears evidence of the universality class of three-state Potts model for these two particular field-driven phase transitions. Two phases emergent above and below the first and fourth critical boundaries can be viewed as two different states of one species (either the singlet-dimer or the polarized triplet state) of different density. On the other hand, the precise nature of field-induced phase transitions inherent to the second and third critical field is more puzzling, because there are strong indications that they should be of the three-state Potts' universality class at higher temperatures and likely of Kosterlitz-Thouless type at lower temperatures.\cite{kinz81,nien84,qian04} Notwithstanding this unclear nature, two domes of continuous field-driven phase transitions separate two period-three quantum phases with a regular alternation of 'singlet-singlet-triplet' or 'singlet-triplet-triplet' dimer states, which manifest themselves in the respective magnetization curves as intermediate one-third and two-thirds magnetization plateaus, respectively. 

\section{Conclusion}
\label{sec:conc}

The present work deals with the magnetization process and low-temperature thermodynamics of the frustrated spin-$\frac{1}{2}$ Heisenberg triangular bilayer, which has been treated by means of various analytical and numerical techniques. The variational method has been adapted in order to find rigorous bounds for the singlet-dimer ground state, while the numerical ED has been used to get exact results for a relatively small ($3 \times 3 \times 2$) finite-size triangular bilayer. Besides, we have developed the localized-magnon approach in order to establish a mapping correspondence with the classical Ising model on a triangular lattice, which has been subsequently analyzed either by exact calculations for small system sizes or by MC simulations for larger system sizes. It should be noticed that a validity of the localized-magnon approach is restricted merely to a highly frustrated parameter region $J_2>3|J_1|$, where the localized many-magnon states determine low-lying part of the energy spectrum.  

Among other matters, it has been demonstrated that the nature of the interdimer interaction $J_1$ fundamentally influences a magnetic behavior of the frustrated spin-$\frac{1}{2}$ Heisenberg triangular bilayer. The FM/AF bilayer with the ferromagnetic interdimer interaction exhibits at low enough temperatures a discontinuous field-driven phase transition accompanied with a finite cusp of the susceptibility, the specific heat and an abrupt  magnetization jump, which gradually diminishes upon increasing temperature until a continuous field-driven phase transition from the Ising universality class is reached at a critical temperature. Contrary to this, the AF/AF bilayer with the antiferromagnetic interdimer interaction displays a sequence of three discontinuous field-driven phase transitions only at zero temperature, which change into four continuous field-driven phase transitions at sufficiently low but nonzero temperatures. Two continuous field-induced transitions closely connected with a breakdown of the singlet-dimer phase and an onset of the saturated ferromagnetic state are from the universality class of three-state Potts model, while another two continuous field-driven phase transition retain this character at higher temperatures and are likely of Kosterlitz-Thouless type at lower temperatures. 

\begin{acknowledgments}
This work was financially supported by the grant of The Ministry of Education, Science, Research and Sport of the Slovak Republic under the contract No. VEGA 1/0043/16 and by the grant of the Slovak Research and Development Agency under the contract No. APVV-16-0186. The work of O.~D. was partially supported by Project FF-30F (No.~0116U001539) from the Ministry of Education and Science of Ukraine.
\end{acknowledgments}

\appendix\section{One-magnon energy spectra}
\label{appa}

\begin{figure}
\begin{center}
\includegraphics[width=0.6\columnwidth]{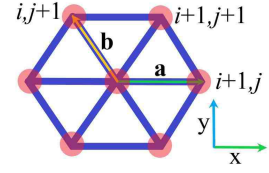}
\end{center}
\vspace{-1.2cm}
\caption{The basis vectors ${\bf{a}}=a_0(1,0)$ and ${\bf{b}}=a_0(-\frac{1}{2},\frac{\sqrt{3}}{2})$ of one triangular layer 
($a_0=1$ is the triangular side length) used for a calculation of the one-magnon energy spectra presented in Appendix~\ref{appa}.}
\label{fig1ap}
\end{figure}

Consider the frustrated spin-$\frac{1}{2}$ Heisenberg model on a triangular bilayer lattice. 
The position of the lattice cells is given by
\begin{eqnarray}
\label{01}
{\bf{R}}&=&m_a{\bf{a}}+m_b{\bf{b}}
=\left(m_a-\frac{m_b}{2}\right)a_0{\bf{i}}+m_b\frac{\sqrt{3}}{2}a_0{\bf{j}},
\nonumber\\
m_a&=&0,1,\ldots,{{L}}-1,
\;\;\;
m_b=0,1,\ldots,{{L}}-1,
\end{eqnarray}
where $a_0$ is the triangle side length, see Fig.~\ref{fig1ap} and Ref.~\onlinecite{krok17}.
The lattice consists of $2N$ sites and $N = L^2$ is the number of cells, i.e. the number of vertical $J_2$ bonds.
The Hamiltonian (\ref{ham}) of the frustrated spin-$\frac{1}{2}$ Heisenberg model on a triangular bilayer 
in the one-magnon subspace reads
\begin{eqnarray}
\label{02}
\hat{\cal H}&=&\sum_{m_a=0}^{{{L}}-1}\sum_{m_b=0}^{{{L}}-1}
\left(
J_2h_{1,m_a,m_b;2,m_a,m_b}
\right.
\nonumber\\
&+& \left. J_1h_{1,m_a,m_b;1,m_a+1,m_b}+J_1h_{2,m_a,m_b;2,m_a+1,m_b}
\right.
\nonumber\\
&+& \left. J_{1}h_{1,m_a,m_b;2,m_a+1,m_b}+J_{1}h_{2,m_a,m_b;1,m_a+1,m_b}
\right.
\nonumber\\
&+& \left. J_1h_{1,m_a,m_b;1,m_a,m_b+1}+J_1h_{2,m_a,m_b;2,m_a,m_b+1}
\right.
\nonumber\\
&+& \left. J_{1}h_{1,m_a,m_b;2,m_a,m_b+1}+J_{1}h_{2,m_a,m_b;1,m_a,m_b+1}
\right.
\nonumber\\
&+& \left. J_1h_{1,m_a,m_b;1,m_a+1,m_b+1}+J_1h_{2,m_a,m_b;2,m_a+1,m_b+1}
\right.
\nonumber\\
&+& \left. J_{1}h_{1,m_a,m_b;2,m_a+1,m_b+1}+J_{1}h_{2,m_a,m_b;1,m_a+1,m_b+1}
\right),
\nonumber\\
h_{i;j}&=&\frac{1}{2}\left(\hat{S}_i^-\hat{S}_j^+ + \hat{S}_j^-\hat{S}_i^+\right)
-\frac{1}{2}\left(\hat{S}_i^-\hat{S}_i^+ + \hat{S}_j^-\hat{S}_j^+\right)+\frac{1}{4}.
\end{eqnarray}

Next, we perform the Fourier transformation:
\begin{eqnarray}
\label{03}
\hat{S}_{l,m_a,m_b}^+
&=&
\frac{1}{{{L}}}\sum_{k_a}\sum_{k_b}\exp\left[{\rm{i}}\left(k_am_a+k_bm_b\right)\right]\hat{S}_{l,{\bf{k}}}^+,
\nonumber\\
\hat{S}_{l,m_a,m_b}^-
&=&
\frac{1}{{{L}}}\sum_{k_a}\sum_{k_b}\exp\left[-{\rm{i}}\left(k_am_a+k_bm_b\right)\right]\hat{S}_{l,{\bf{k}}}^-,
\nonumber\\
l&=&1,2,
\nonumber\\
k_a&=&\frac{2\pi}{{{L}}} z_a, z_a=0,1,\ldots,{{L}}-1,
\nonumber\\
k_b&=&\frac{2\pi}{{{L}}} z_b, z_b=0,1,\ldots,{{L}}-1,
\nonumber\\
{\bf{k}}&=&\frac{k_a}{a_0}{\bf{i}}+\frac{k_a+2k_b}{\sqrt{3}a_0}{\bf{j}}.
\end{eqnarray}
Clearly,
\begin{eqnarray}
\label{04}
&&
\sum_{m_a=0}^{{{L}}-1}\sum_{m_b=0}^{{{L}}-1} J_2h_{1,m_a,m_b;2,m_a,m_b}
\nonumber\\
&=&
\sum_{\bf{k}}
\left[
\frac{J_2}{2}\left(\hat{S}^-_{1,{\bf{k}}} \hat{S}^+_{2,{\bf{k}}} + \hat{S}^-_{2,{\bf{k}}} \hat{S}^+_{1,{\bf{k}}}\right)
\right.
\nonumber\\
&-& \left. \frac{J_2}{2}\left(\hat{S}^-_{1,{\bf{k}}} \hat{S}^+_{1,{\bf{k}}} + \hat{S}^-_{2,{\bf{k}}} \hat{S}^+_{2,{\bf{k}}}\right)
\right] +{{N}}\frac{J_2}{4},
\end{eqnarray}
\begin{eqnarray}
\label{05}
\sum_{m_a=0}^{{{L}}-1}\sum_{m_b=0}^{{{L}}-1}
\left(
J_1h_{1,m_a,m_b;1,m_a+1,m_b}+J_1h_{2,m_a,m_b;2,m_a+1,m_b}
\right.
\nonumber\\
\left.
+J_{1}h_{1,m_a,m_b;2,m_a+1,m_b}+J_{1}h_{2,m_a,m_b;1,m_a+1,m_b}
\right)
\nonumber\\
=
\sum_{\bf{k}}
\left[
J_1 \left(\cos k_a -2\right)
\left(\hat{S}^-_{1,{\bf{k}}} \hat{S}^+_{1,{\bf{k}}} + \hat{S}^-_{2,{\bf{k}}} \hat{S}^+_{2,{\bf{k}}}\right)
\right.
\nonumber\\
\left.
+J_{1}\cos k_a
\left(\hat{S}^-_{1,{\bf{k}}} \hat{S}^+_{2,{\bf{k}}} + \hat{S}^-_{2,{\bf{k}}} \hat{S}^+_{1,{\bf{k}}}\right)
\right] + N J_1
\nonumber\\
\end{eqnarray}
etc.
Therefore the Hamiltonian can be cast into
\begin{eqnarray}
\label{06}
\hat{\cal H}=\sum_{{\bf{k}}}
\left(
\begin{array}{cc}
\hat{S}^-_{1,{\bf{k}}} & \hat{S}^-_{2,{\bf{k}}}
\end{array}
\right)
\left(
\begin{array}{cc}
H_{11} & H_{12}\\
H_{21} & H_{22}
\end{array}
\right)
\left(
\begin{array}{c}
\hat{S}^+_{1,{\bf{k}}} \\
\hat{S}^+_{2,{\bf{k}}}
\end{array}
\right)
\nonumber\\
+{{N}}\left(\frac{J_2}{4}+ 3 J_1 \right),
\nonumber\\
H_{11}=H_{22}= J_1\left[\cos k_a+\cos k_b+\cos\left(k_a+k_b\right)\right]
\nonumber\\
-6 J_1 -\frac{J_2}{2},
\nonumber\\
H_{12}=H_{21}= J_{1}\left[\cos k_a+\cos k_b+\cos\left(k_a+k_b\right)\right]
\nonumber\\
+\frac{J_2}{2}.
\end{eqnarray}
One-magnon energies $\varepsilon_{\bf{k}}^{(1,2)}=H_{11}\mp H_{12}$
are as follows:
\begin{eqnarray}
\label{08}
\varepsilon_{\bf{k}}^{(1)}
&=&-J_2-6J_1,
\nonumber\\
\varepsilon_{\bf{k}}^{(2)}
&=&2J_1\left[\cos k_a+\cos k_b+\cos\left(k_a+k_b\right)-3\right]
\nonumber\\
&=&8J_1\left(\cos \frac{k_a}{2}\cos \frac{k_b}{2}\cos\frac{k_a+k_b}{2}-1 \right).
\end{eqnarray}

\section{Exact results for the effective $3 \times 3$ triangular Ising model}
\label{appb}

\begin{figure}
\begin{center}
\includegraphics[width=0.20\textwidth]{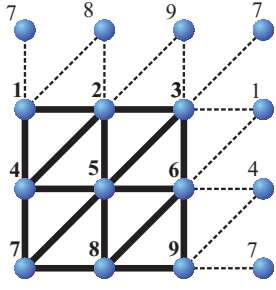}
\end{center}
\vspace{-0.3cm}
\caption{A schematic illustration of the effective $3 \times 3$ triangular Ising model under the periodic boundary conditions.}
\label{system2}
\end{figure}

\begin{figure}
\begin{center}
\includegraphics[width=0.20\textwidth]{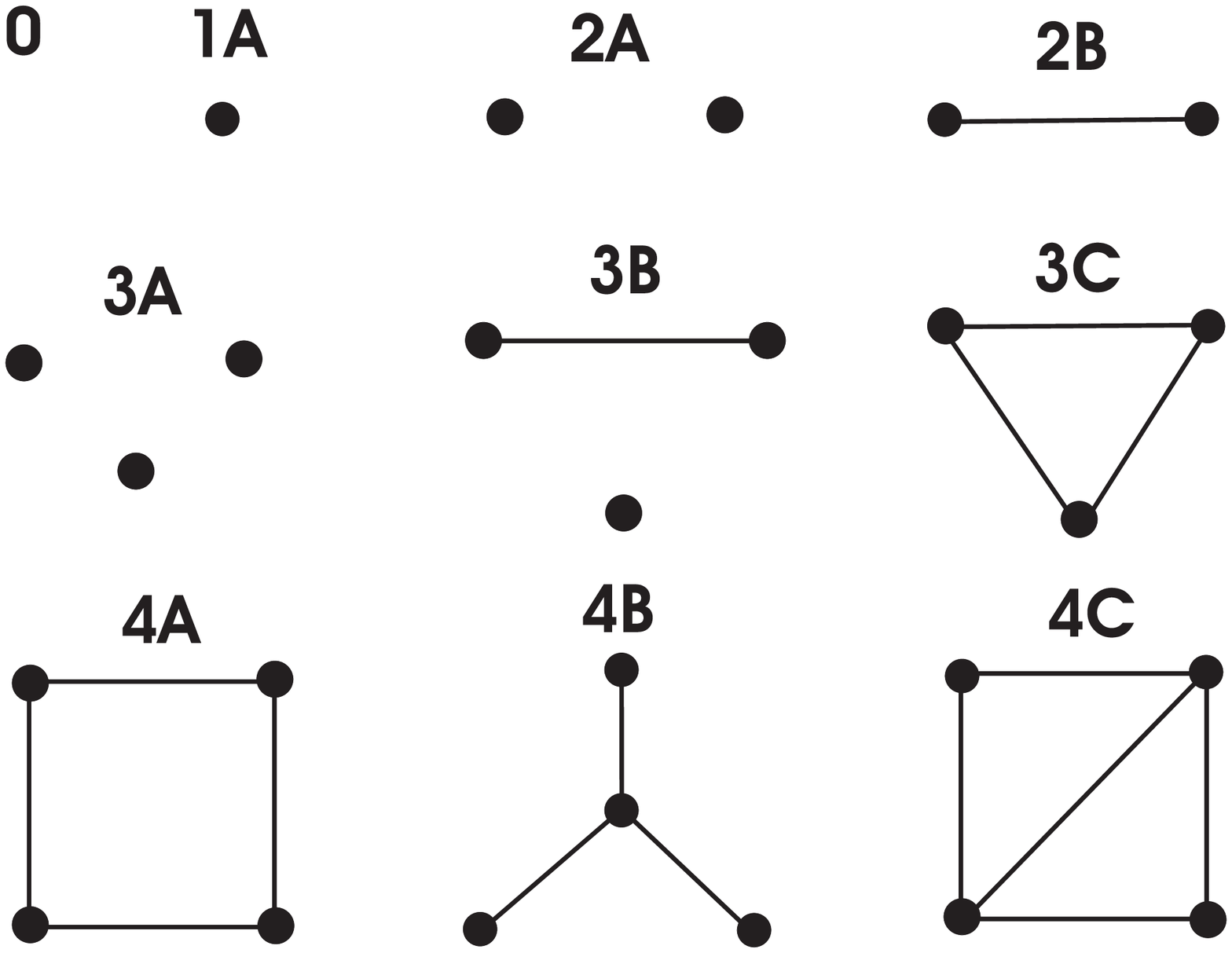}
\end{center}
\vspace{-0.4cm}
\caption{A schematic illustration of the induced subgraphs, which correspond to all possible spin configurations 
of the effective $3 \times 3$ triangular Ising model.}
\label{podgrafy}
\end{figure}

In this part we will adapt the graph-theoretical approach developed in Ref. \onlinecite{stre15} in order to find an exact solution of the effective $3 \times 3$ triangular Ising model given by the Hamiltonian (\ref{hamef}) with the specific value of linear size $L=3$ (see
Fig.~\ref{system2} for a schematic illustration). It  should be mentioned that each individual spin configuration can be represented according to Ref. \onlinecite{stre15} by an induced subgraph and the overall energy can be calculated from the formula 
\begin{equation}
E = J_{\rm eff} (27 - 2 \tilde{d}_t) - h_{\rm eff} S_T,
\label{enc}
\end{equation}
where $\tilde{d}_t$ determines the total number of unlike oriented adjacent spin pairs and $S_T$ represents the total spin for a given spin configuration. Note that the total spin $S_T = 9 - 2n_t$ can be related to the total number of vertices $n_t$ within a given induced subgraph and $\tilde{d}_t$ determines the sum of their complementary degrees. The induced subgraphs corresponding to all available  spin configurations of the effective $3 \times 3$ triangular Ising model are listed in Table \ref{tab_podgrafy} and schematically illustrated in Fig. \ref{podgrafy}. 

\begin{table}
\vspace{-0.5cm}
\caption{Spin configurations of the effective $3 \times 3$ triangular Ising model classified according to the total spin $S_{T} \geq 0$, the total number of flipped Ising spins ($n_t$), the degeneracy (deg), the total number of unlike oriented adjacent spin pairs ($\tilde{d}_t$), the overall energy and the corresponding induced subgraph (see Fig.~\ref{podgrafy} for schematic illustration of induced subgraphs). The values $n_t$ and $\tilde{d}_t$ coincide with the total number of vertices in a given subgraph and the sum of their complementary degrees, respectively.}
\vspace{-0.2cm}
\begin{center}
\begin{tabular}{|c|c|c|c|c|c|}
\hline $ S_{T} $ & $ n_t $ & deg & $ \tilde{d}_t $ & energy & subgraph \\  
\hline 9 & 0 & 1 & 0 & $27J_{\rm eff}-9h_{\rm eff}$ & 0 \\ 
\hline 7 & 1 & 9 & 6 & $15J_{\rm eff}-7h_{\rm eff}$ & 1A \\ 
\hline 5 & 2 & 9 & 12 & $3J_{\rm eff}-5h_{\rm eff}$ & 2A \\ 
\hline 5 & 2 & 27 & 10 & $7J_{\rm eff}-5h_{\rm eff} $ & 2B \\ 
\hline 3 & 3 & 3 & 18 & $-9J_{\rm eff}-3h_{\rm eff} $ & 3A \\ 
\hline 3 & 3 & 54 & 14 & $-J_{\rm eff}-3h_{\rm eff}$ & 3B \\ 
\hline 3 & 3 & 27 & 12 & $3J_{\rm eff}-3h_{\rm eff}$& 3C \\ 
\hline 1 & 4 & 27 & 16 & $-5J_{\rm eff}-h_{\rm eff}$ & 4A \\ 
\hline 1 & 4 & 81 & 14 & $-J_{\rm eff}-h_{\rm eff} $ & 4B \\ 
\hline 1 & 4 & 18 & 18 & $-9J_{\rm eff}-h_{\rm eff} $ & 4C \\ 
\hline 
\end{tabular} 
\end{center}
\label{tab_podgrafy}
\end{table}
A summation over the overall energy spectrum affords the following exact result for the partition function of the effective $3 \times 3$ triangular Ising model 
\begin{widetext}
\begin{eqnarray}
Z \!&=&\! 2\exp\left(-27\beta J_{\rm eff}\right)\cosh(9\beta h_{\rm eff}) + 18\exp\left(-15\beta J_{\rm eff}\right)\cosh(7\beta h_{\rm eff}) + 18\exp\left(-3\beta J_{\rm eff}\right)\cosh(5\beta h_{\rm eff}) \nonumber \\ 
\!&+&\! 54\exp\left(-7\beta J_{\rm eff}\right)\cosh(5\beta h_{\rm eff})  + 6\exp\left(9\beta J_{\rm eff}\right)\cosh(3\beta h_{\rm eff}) + 108\exp\left(\beta J_{\rm eff}\right)\cosh(3\beta h_{\rm eff})   \label{PFeffective} \\ 
\!&+&\! 54\exp\left(-3\beta J_{\rm eff}\right)\cosh(3\beta h_{\rm eff}) + 54\exp\left(5\beta J_{\rm eff}\right)\cosh(\beta h_{\rm eff}) + 36\exp\left(9\beta J_{\rm eff}\right)\cosh(\beta h_{\rm eff}) +	162\exp\left(\beta J_{\rm eff}\right)\cosh(\beta h_{\rm eff}). \nonumber 
\end{eqnarray}
\end{widetext}
The exact result (\ref{PFeffective}) can be straightforwardly used for a calculation of the free energy, magnetization, susceptibility and specific heat by standard means.

\end{document}